\documentclass[aps,prd,nofootinbib,floatfix,superscriptaddress,secnumarabic]{revtex4}
\usepackage{bm}
\usepackage{graphicx}%
\usepackage{epsf}
\usepackage{epsfig}

 \usepackage{amsmath}
\usepackage{amsfonts,amssymb}

\usepackage{colordvi}
\usepackage{color}
\usepackage{lscape}
\usepackage{rotfloat}
\usepackage{rotating}

\usepackage{dsfont}
\usepackage{slashed}%
\usepackage{booktabs}

\newcommand{\be}{\begin{equation}}
\newcommand{\ee}{\end{equation}}
\newcommand{\bea}{\begin{eqnarray}}
\newcommand{\eea}{\end{eqnarray}}

\newcommand{\MSbar}{{\overline{\rm MS}}}
\newcommand{\amp}{{\rm amp}}

\newcommand{\qslash}{{\not{\hspace{-0.05cm}q}}}

\newcommand{\la}{\lambda}

\begin{document}

\title{Supersymmetric QCD: Renormalization and Mixing of Composite Operators}

\author{Marios Costa} 
\email{kosta.marios@ucy.ac.cy}
\affiliation{$\it{Department \ of \ Physics, \ University \ of \ Cyprus, \ POB \ 20537, \ 1678, \ Nicosia, \ Cyprus}$ \vspace{0.5cm}}
\author{Haralambos Panagopoulos} 
\email{haris@ucy.ac.cy}
\affiliation{$\it{Department \ of \ Physics, \ University \ of \ Cyprus, \ POB \ 20537, \ 1678, \ Nicosia, \ Cyprus}$ \vspace{0.5cm}}

\begin{abstract}
We study $4$-dimensional SQCD with gauge group $SU(N_c)$ and $N_f$ flavors of chiral super-multiplets on the lattice. We perform extensive calculations of matrix elements and renormalization factors of composite operators in Perturbation Theory. In particular, we compute the renormalization factors of quark and squark bilinears, as well as their mixing at the quantum level with gluino and gluon bilinear operators. From these results we construct correctly renormalized composite operators, which are free of mixing effects and may be employed in non-perturbative studies of Supersymmetry. All our calculations have been performed with massive matter fields, in order to regulate the infrared singularities which are inherent in renormalizing squark bilinears. Furthermore, the quark and squark propagators are computed in momentum space with nonzero masses.

This work is a feasibility study for lattice computations relevant to a number of observables, such as spectra and distribution functions of hadrons, but in the context of supersymmetric QCD, as a forerunner to lattice investigations of SUSY extensions of the Standard Model. 
\end{abstract}

\maketitle

\section{Introduction}
\bigskip
Current intensive searches for Physics Beyond the Standard Model (BSM) are becoming a very timely endeavor \cite{Tanabashi:2018}, given the precision experiments at the Large Hadron Collider and elsewhere; at the same time, numerical studies of BSM Physics are more viable due to the advent of lattice formulations which preserve chiral symmetry \cite{Kaplan:1992bt, Neuberger:1997fp, Luscher:1998}. In particular, the study of supersymmetric models on the lattice, which began a long time ago~\cite{Curci:1986sm}, has been an object of intense research activity in recent years \cite{Kaplan:2009, Catterall:2014vga, Joseph:2015xwa, Ali:2018fbq, Endrodi:2018ikq}, and applications to supersymmetric extensions of the Standard Model are gradually becoming within reach. Studies of hadronic properties using the lattice formulation of Supersymmetric Quantum Chromodynamics (SQCD) rely on the computation of matrix elements and correlation functions of composite operators, made out of quark ($\psi$), gluino ($\lambda$), gluon ($u$), squark ($A$) fields. These operators are of great phenomenological interest in the non-supersymmetric case, since they are employed in the calculation of certain transition amplitudes among bound states of particles and in the extraction of meson and baryon form factors. Correlation functions of such operators calculated in lattice SQCD therefore provide interesting probes of physical properties of the theory. A proper renormalization of these operators is most often indispensable for the extraction of results from the lattice. The main objective of this work is the calculation of the quantum corrections to a complete basis of ``ultra local'' bilinear currents, using both dimensional regularization and lattice regularization. We consider both flavor singlet and nonsinglet operators.  

Within the SQCD formulation we compute the quark, squark propagators and all 2-pt Green's functions of bilinear operators, made out of quark and squark fields. Our computations are performed to one loop and to lowest order in the lattice spacing, $a$; also, in order to avoid potential infrared singularities in Green's functions of squark bilinears, we have employed massive chiral supermultiplets throughout. We extract from the above quantities the renormalization factors of the quark and squark fields and masses. Quantum corrections induce mixing of some of the bilinear operators which we study, both among themselves and with gluon and gluino bilinears having the same quantum numbers; we compute all the corresponding mixing coefficients, in $\MSbar$ renormalization scheme; the values of these coefficients can also be readily derived from the renormalized Green's functions which we provide. Our calculations also reproduce the  Adler-Bell-Jackiw (ABJ) anomaly of the axial vector quark current to one-loop order.

This work is a continuation to our recent paper~\cite{MC:2018}, in which we presented our perturbative results for the renormalization factors of the coupling constant ($Z_g$) and of the quark ($Z_\psi$), gluon ($Z_u$), gluino ($Z_\lambda$), squark ($Z_{A_\pm}$), and ghost ($Z_c$) fields in the continuum and on the lattice; they were the first one-loop computations of these quantities using a lattice discretization of the action of SQCD. 

The paper is organized as follows: Sec.~\ref{sec2} contains all relevant definitions and the calculational setup. In Sec.~\ref{sec3} we present our computation and results both for dimensional and lattice regularizations. Finally, we conclude in Sec.~\ref{summary} with a summary and a discussion of our results and possible future extensions of our work.

\section{Formulation and Computational Setup}
\label{sec2}

\subsection{Lattice Action}
\label{sec2.1}

In our lattice calculation, we extend Wilson's formulation of the QCD action, to encompass SUSY partner fields as well. In this standard discretization quarks, squarks and gluinos live on the lattice sites, and gluons live on the links of the lattice: $U_\mu (x) = e^{i g a T^{\alpha} u_\mu^\alpha (x+a\hat{\mu}/2)}$; $\alpha$ is a color index in the adjoint representation of the gauge group. This formulation leaves no SUSY generators intact, and it also breaks chiral symmetry; it thus represents a ``worst case'' scenario, which is worth investigating in order to address the complications \cite{Giedt} which will arise in numerical simulations of SUSY theories. In our ongoing investigation we plan to address also improved actions, so that we can check to what extent some of the SUSY breaking effects can be alleviated. For Wilson-type quarks ($\psi$) and gluinos ($\lambda$), the Euclidean action ${\cal S}^{L}_{\rm SQCD}$ on the lattice becomes ($A_\pm$ are the squark field components):       
\bea
{\cal S}^{L}_{\rm SQCD} & = & a^4 \sum_x \Big[ \frac{N_c}{g^2} \sum_{\mu,\,\nu}\left(1-\frac{1}{N_c}\, {\rm Tr} U_{\mu\,\nu} \right ) + \sum_{\mu} {\rm Tr} \left(\bar \lambda_M \gamma_\mu {\cal{D}}_\mu\lambda_M \right ) - a \frac{r}{2} {\rm Tr}\left(\bar \lambda_M  {\cal{D}}^2 \lambda_M \right) \nonumber \\ 
&+&\sum_{\mu}\left( {\cal{D}}_\mu A_+^{\dagger}{\cal{D}}_\mu A_+ + {\cal{D}}_\mu A_- {\cal{D}}_\mu A_-^{\dagger}+ \bar \psi_D \gamma_\mu {\cal{D}}_\mu \psi_D \right) - a \frac{r}{2} \bar \psi_D  {\cal{D}}^2 \psi_D \nonumber \\
&+&i \sqrt2 g \big( A^{\dagger}_+ \bar{\lambda}^{\alpha}_M T^{\alpha} P_+ \psi_D  -  \bar{\psi}_D P_- \lambda^{\alpha}_M  T^{\alpha} A_+ +  A_- \bar{\lambda}^{\alpha}_M T^{\alpha} P_- \psi_D  -  \bar{\psi}_D P_+ \lambda^{\alpha}_M  T^{\alpha} A_-^{\dagger}\big)\nonumber\\  
&+& \frac{1}{2} g^2 (A^{\dagger}_+ T^{\alpha} A_+ -  A_- T^{\alpha} A^{\dagger}_-)^2 - m ( \bar \psi_D \psi_D - m A^{\dagger}_+ A_+  - m A_- A^{\dagger}_-)\Big] \,,
\label{susylagrLattice}
\eea
where: $U_{\mu \nu}(x) =U_\mu(x)U_\nu(x+a\hat\mu)U^\dagger_\mu(x+a\hat\nu)U_\nu^\dagger(x)$, and a summation over flavors is understood in the last three lines of Eq.~(\ref{susylagrLattice}). The 4-vector $x$ is restricted to the values $x = na$, with $n$ being an integer 4-vector. The terms proportional to the Wilson parameter, $r$, eliminate the problem of fermion doubling, at the expense of breaking chiral invariance. In the limit $a \to 0$ the lattice action reproduces the continuum one.

The definitions of the covariant derivatives are as follows:
\bea
{\cal{D}}_\mu\lambda_M(x) &\equiv& \frac{1}{2a} \Big[ U_\mu (x) \lambda_M (x + a \hat{\mu}) U_\mu^\dagger (x) - U_\mu^\dagger (x - a \hat{\mu}) \lambda_M (x - a \hat{\mu}) U_\mu(x - a \hat{\mu}) \Big] \\
{\cal D}^2 \lambda_M(x) &\equiv& \frac{1}{a^2} \sum_\mu \Big[ U_\mu (x)  \lambda_M (x + a \hat{\mu}) U_\mu^\dagger (x)  - 2 \lambda_M(x) +  U_\mu^\dagger (x - a \hat{\mu}) \lambda_M (x - a \hat{\mu}) U_\mu(x - a \hat{\mu})\Big]\\
{\cal{D}}_\mu \psi_D(x) &\equiv& \frac{1}{2a}\Big[ U_\mu (x) \psi_D (x + a \hat{\mu})  - U_\mu^\dagger (x - a \hat{\mu}) \psi_D (x - a \hat{\mu})\Big]\\  
{\cal D}^2 \psi_D(x) &\equiv& \frac{1}{a^2} \sum_\mu \Big[U_\mu (x) \psi_D (x + a \hat{\mu})  - 2 \psi_D(x) +  U_\mu^\dagger (x - a \hat{\mu}) \psi_D (x - a \hat{\mu})\Big]\\
{\cal{D}}_\mu A_+(x) &\equiv& \frac{1}{a} \Big[  U_\mu (x) A_+(x + a \hat{\mu}) - A_+(x)   \Big]\\
{\cal{D}}_\mu A_+^{\dagger}(x) &\equiv& \frac{1}{a} \Big[A_+^{\dagger}(x + a \hat{\mu}) U_\mu^{\dagger}(x)  -  A_+^\dagger(x)\Big]\\
{\cal{D}}_\mu A_-(x) &\equiv& \frac{1}{a} \Big[A_-(x + a \hat{\mu}) U_\mu^{\dagger}(x)  -  A_-(x)\Big]\\
{\cal{D}}_\mu A_-^{\dagger}(x) &\equiv& \frac{1}{a} \Big[U_\mu (x) A_-^{\dagger}(x + a \hat{\mu})   -  A_-^{\dagger}(x) \Big]
\eea
A gauge-fixing term, together with the compensating ghost field term, must be added to the action, in order to avoid divergences from the  integration over gauge orbits; these terms are the same as in the non-supersymmetric case. Similarly, a standard ``measure'' term must be added to the action, in order to accound for the Jacobian in the change of integration variables: $U_\mu \to u_\mu$\,. All the details of the continuum and the lattice actions can be found in Ref.\cite{MC:2018}.

\subsection{Bilinear operators and their mixing}
\label{sec2.2}

In studying the properties of physical states, the main observables are Green's functions of operators made of quark fields, having the form ${\cal O}_i^\psi(x) = \bar \psi(x) \Gamma_i \psi(x)$, where $\Gamma_i$ denotes all possible distinct products of Dirac matrices, as well as operators made of squark fields ${\cal O}^A(x) = A^\dagger(x) A(x)$, along with operators of higher dimensionality. The matter fields are considered to be massive, and for completeness' sake we calculate the quark and squark propagators with nonzero masses; in this way, we have control over infrared (IR) divergences. Ultraviolet (UV) divergences are treated by a standard regularization, either the lattice (L) or dimensional regularization (DR). 

The first new quantities of this paper are the renormalization factors for the squark and quark masses in the $\MSbar$ and $RI'$ schemes. After these computations, we focus on the matrix elements of composite bilinear operators. As we will see below, some of the operators carrying the same quantum numbers mix together beyond tree level. In order to determine their mixing coefficients, we calculate certain 2-pt Green's functions of these operators. More specifically, we calculate the 2-pt Green's function of squark bilinears with external squarks and gluons, as well as the 2-pt Green's functions of quark bilinears with one external quark-antiquark pair, or one gluino-antigluino pair, or two gluons, or two squarks. All of our results are computed as functions of the coupling constant, the number of colors, the gauge fixing parameter and the external momentum. The renormalization conditions which we will impose involve the renormalization factors of the fields, that we have computed in Ref.\cite{MC:2018}.

We identified all operators which can possibly mix with ${\cal O}^\psi_{i}$ and all Green's functions, e.g. $\langle  \la(x) {\cal O}^\psi_i(z) \bar \la(y) \rangle$, which must be calculated in order to compute those elements of the mixing matrix which are relevant for the renormalization of ${\cal O}^\psi_{i}$ and ${\cal O}^A$. 

The bilinear operators could in principle mix with four types of operators having the same quantum numbers. The four types are as follows: Type I are gauge invariant operators. Type II are operators which are not gauge invariant but are the BRST variation \cite{Suzuki} of some other operators. Type III operators vanish by the  equations of motion. Type IV are operators which are not linear combinations of type I, II and III. By general renormalization theorems (see, e.g., Ref.~\cite{Collins:1984xc}), type I operators will not mix with type IV operators. We list the type I operators in Table~\ref{tb:non-singlet} for the flavor non-singlet case ($\bar \psi \Gamma_i \psi \equiv \bar \psi_f \Gamma_i \psi_{f'}$) and in Table~\ref{tb:singlet} additional operators which show up in the flavor singlet case ($\bar \psi \Gamma_i \psi \equiv \frac{1}{N_f}\sum_f \bar \psi_f \Gamma_i \psi_f$). Different values of  the index ``$i$'' lead to the following possibilities for the Dirac matrices: (scalar) $\Gamma_S= 1$, (pseudoscalar) $\Gamma_P=\gamma_5$, (vector) $\Gamma_V =\gamma_\mu$, (axial vector) $\Gamma_{AV}=\gamma_5\gamma_\mu$ and tensor $\Gamma_T= [\gamma_{\mu},\gamma_{\nu}]/2$. In Tables~\ref{tb:non-singlet} and \ref{tb:singlet} we also include operators with lower dimensionalities, even though they do not mix with quark bilinears in dimensional regularization; they do however show up in the lattice formulation.  Indeed, on the lattice, the number of operators which mix among themselves is considerably greater than in the continuum regularization. The perturbative computations of all relevant Green's functions of operators ${\cal O}^\psi_{i}$ and ${\cal O}^A$  will be followed by the construction of the mixing matrix, which may also involve non gauge invariant (but BRST invariant) operators or operators which vanish by the equations of motion.

In this work we concentrate on extracting the mixing coefficients between quark, squark, gluino and gluon bilinears, which are relevant for physical external states and thus we do not take into account the ghost bilinears. As it turns out, flavor-singlet quark bilinears mix with gluino bilinears: ${\cal O}_i^\lambda(x) = {\rm Tr}\left(\bar \la(x) \Gamma_i \la(x)\right)$. A particularly rich mixing pattern emerging from Tables~\ref{tb:non-singlet} and \ref{tb:singlet} regards the case of scalar, pseudoscalar and vector quark bilinears, since they can mix with a variety of gluon and squark bilinear operators. Note also that the scalar operators mix with the identity at the quantum level. 

In order to make the BRST symmetry explicit and elucidate the mixing with non-gauge-invariant operators, we write the Faddeev-Popov action, using a new auxiliary field $B^\alpha$:
\begin{equation}
 S_{FP}= \int d^4x\left [\frac{\alpha}{2}(B^\alpha)^2 - B^{\alpha} \partial^\mu u_\mu^\alpha-\bar{c}^\alpha \partial^{\mu}D^{\alpha \beta}_\mu  c^\beta \right ]. 
\label{sgf}
\end{equation}
Under BRST transformations, the fields appearing in $S_{FP}$ behave as follows:
\begin{eqnarray}
u^{\alpha}_\mu &\rightarrow & u^\alpha_\mu + D_\mu^{\alpha\beta}c^\beta \ \xi ,\nonumber \\
c^\alpha &\rightarrow & c^\alpha -\frac{g}{2}f^{\alpha\beta\gamma}c^\beta c^\gamma \ \xi, \nonumber \\
\bar{c}^\alpha &\rightarrow & \bar{c}^\alpha +B^\alpha \ \xi,\nonumber \\
B^\alpha &\rightarrow & B^\alpha,
\label{ffb}
\end{eqnarray}
where $\xi$  is an infinitesimal anticommuting parameter. Under these transformations, $S_{FP}$ is indeed invariant. Given that the effect of a BRST transformation on gauge and matter fields is that of a gauge transformation (with anticommuting parameter), all other parts of the action, being gauge invariant, will automatically also be BRST invariant.

From Eqs.~(\ref{ffb}), we see that type II operators may mix in the flavor-singlet case, e.g. :
\be
\delta_{BRST}\left(u_\mu^{\alpha} \bar c^{\alpha}\right) = \left(u_\mu^{\alpha} B^{\alpha} + \bar c^\alpha \, D_\mu^{\alpha \beta} c^\beta \right) \Rightarrow \delta_{BRST}\left(u_\mu^{\alpha} B^{\alpha} + \bar c^\alpha \, D_\mu^{\alpha \beta} c^\beta \right) = 0
\label{NGinv}
\ee
This potential mixing of the non-gauge-invariant operator $u_\mu B + \bar c D_\mu c = (u_\mu \partial_\nu u^\nu)/\alpha +\bar c D_\mu c $ (which is also present in the non-supersymmetric case), is inconsequential if one is interested in physical external states with transverse polarization. Other flavor non-singlet operators, such as $A_+^{\dagger}\partial_\mu A_+$, etc., are not of types I, II or III and therefore cannot mix with gauge invariant operators. The presence of a global $U(1)$ symmetry, which is preserved by the SQCD action, both in the continuum and on the lattice, forbids 3-squark operators from mixing with quark bilinears.

By investigating the behavior of dimension-2 and -3 bilinear operators under parity, $\cal{P}$, and charge conjugation, $\cal{C}$, we group them to the categories $S$, $P$, $V$, $A$ and $T$ in Tables~\ref{tb:non-singlet} and \ref{tb:singlet}, based on the trasformation properties of the quark bilinears. These are symmetries of the action and their definitions are presented below.
\be
{\cal{P}}:\left \{\begin{array}{ll}
&\hspace{-.3cm} U_0(x)\rightarrow U_0(x_{\cal{P}})\, ,\qquad U_k(x)\rightarrow U_k^{\dagger}(x_{\cal{P}}-a\hat{k})\, ,\qquad k=1,2,3\\
&\hspace{-.3cm} \psi_f(x)\rightarrow \gamma_0  \psi_f(x_{\cal{P}})\\
&\hspace{-.3cm}\bar{ \psi}_f(x) \rightarrow\bar{ \psi}_f(x_{\cal{P}})\gamma_0\\
&\hspace{-.3cm} \la_f(x)\rightarrow \gamma_0  \la_f(x_{\cal{P}})\\
&\hspace{-.3cm}\bar{ \la}_f(x) \rightarrow\bar{ \la}_f(x_{\cal{P}})\gamma_0\\
&\hspace{-.3cm} A_\pm(x) \rightarrow A_\mp^\dagger(x_{\cal{P}})\\
&\hspace{-.3cm} A_\pm^\dagger(x) \rightarrow A_\mp(x_{\cal{P}})
\end{array}\right .
\label{Parity}
\ee
where $x_{\cal{P}}=(-{\bf{x}},x_0)$.

\be
{\mathcal {C}}:\left \{\begin{array}{ll}
&\hspace{-.3cm}U_\mu(x)\rightarrow U_\mu^{\star}(x)\, ,\quad \mu=0,1,2,3\\
&\hspace{-.3cm}\psi(x)\rightarrow i\gamma_0 \gamma_2 \bar{ \psi}(x)^{T}\\
&\hspace{-.3cm}\bar{\psi}(x)\rightarrow-{\psi}(x)^{T}i\gamma_0\gamma_2\\
&\hspace{-.3cm} \la(x) \rightarrow -i\gamma_0 \gamma_2 \bar{\la}(x)^{T}\\
&\hspace{-.3cm}\bar{\la}(x) \rightarrow {\la}(x)^{T}i\gamma_0\gamma_2\\
&\hspace{-.3cm}A_\pm(x) \rightarrow A_\mp(x) \\
&\hspace{-.3cm}A_\pm^\dagger(x) \rightarrow A_\mp^\dagger(x)
\end{array}\right . 
\label{Chargeconjugation}
\ee
where $^{\,T}$ means transpose. For bilinear operators, it is convenient to define a new trasformation ${\cal C'}$, which is a combination of ${\cal C}$ with an exchange in the flavors of the two fields and in their respective masses; this transformation is a (spurionic) symmetry of the action. The operators shown in Table~\ref{tb:non-singlet} are eigenstates of both ${\cal P}$ and ${\cal C'}$. Mixing with further operators, such as $A_+^{\dagger} A_+-A_- A_-^{\dagger}$ and $A_+^{\dagger} D_\mu A_-^{\dagger} - A_- D_\mu A_+$, is not allowed, due to incompatible eigenvalues under $\cal{C'}$.

\begin{table}[ht]
\begin{center}

\begin{tabular}{c|c|c|c}
\hline
\hline
\textbf{Operators}  & \,\,\,\,\textbf{$\cal{P}$}\,\,\,\, & \,\,\,\,\textbf{$\cal{C'}$}\,\,\,\, &\textbf{Category}  \\ [0.5ex] \hline
$\bar \psi \psi$& $+$& $+$&$S$  \\[0.5ex]\hline
$\bar \psi \gamma_5 \psi$& $-$ &+ &$P$  \\[0.5ex]\hline
$\bar \psi \gamma_\mu \psi$&$(-1)^\mu$ & $-$ & $V$ \\[0.5ex]\hline
$\bar \psi \gamma_5 \gamma_\mu \psi$&$-(-1)^\mu$ &$+$&$AV$  \\[0.5ex]\hline
$\frac{1}{2}\bar \psi [\gamma_\mu , \gamma_\nu] \psi$ &$(-1)^\mu(-1)^\nu$ &$-$ &$T$  \\[0.5ex]\hline
$A_+^{\dagger} A_++A_- A_-^{\dagger}$&$+$&$+$&$S$  \\[0.5ex]\hline
$A_+^{\dagger} A_-^{\dagger} + A_- A_+$&$+$&$+$& $S$  \\[0.5ex]\hline
$A_+^{\dagger} A_-^{\dagger} - A_- A_+$&$-$&$+$& $P$ \\[0.5ex]\hline
$(m_f + m_{f'}) (A_+^{\dagger} A_+ + A_- A_-^{\dagger})$&$+$&$+$& $S$  \\[0.5ex]\hline
$(m_f - m_{f'}) (A_+^{\dagger} A_+ - A_- A_-^{\dagger})$&$-$&$+$& $P$  \\[0.5ex]\hline
$(m_f + m_{f'}) (A_+^{\dagger} A_-^{\dagger} + A_- A_+)$&$+$&$+$& $S$  \\[0.5ex]\hline
$(m_f + m_{f'}) (A_+^{\dagger} A_-^{\dagger} - A_- A_+)$&$-$&$+$& $P$  \\[0.5ex]\hline
$A_+^{\dagger} D_\mu A_+ + A_- D_\mu  A_-^{\dagger}$&$(-1)^\mu$&$-$& $V$  \\[0.5ex]\hline
$A_+^{\dagger} D_\mu A_+ - A_- D_\mu  A_-^{\dagger}$&$-(-1)^\mu$&$+$& $AV$  \\[0.5ex]\hline
$A_+^{\dagger} D_\mu A_-^{\dagger} + A_- D_\mu A_+$&$(-1)^\mu$&$-$&$V$ \\[0.5ex]\hline
\hline
\end{tabular}
\caption{Quark bilinears and other operators with which these can mix, in the flavor non-singlet case. Only gauge invariant operators appear in this case. Operators with lower dimensionality will mix on the lattice. All operators appearing in this table are eigenstates of $\cal{P}$ and $\cal{C'}$. In the above operators, the matter fields should be explicitly identified with a flavor index. The flavor indices carried by the left fields ($f$) differ from those of right fields ($f'$). The shorthand $(-1)^\mu$ means $+$ for $\mu=0$ and $-$ for $\mu= 1, 2, 3$.}
\label{tb:non-singlet}
\end{center}
\end{table}

\begin{table}[ht]
\begin{center}
\begin{tabular}{ c| c} 
\hline
\hline
\textbf{Operators}  & \textbf{Category}  \\ [0.5ex] \hline
$\openone$& $S$  \\[0.5ex]\hline
${\rm Tr}\left(\bar \la \la\right)$& $S$  \\[0.5ex]\hline
${\rm Tr}\left(\bar \la \gamma_5 \la\right)$& $P$  \\[0.5ex]\hline
${\rm Tr}\left(\bar \la \gamma_\mu \la\right)$& $V$ \\[0.5ex]\hline
${\rm Tr}\left(\bar \la \gamma_5 \gamma_\mu \la\right)$& $AV$  \\[0.5ex]\hline
${\rm Tr}\left(\frac{1}{2}\bar \la [\gamma_\mu , \gamma_\nu] \la\right)$ & $T$  \\[0.5ex]\hline
\hline
\end{tabular}
\caption{Additional operators which can mix with quark bilinears in the flavor singlet case. A non-gauge invariant (but BRST invariant) operator which can mix is shown in Eq.~(\ref{NGinv}). There are also mixings with other linear combinations of squark bilinears coming from parity and charge conjugation.}
\label{tb:singlet}
\end{center}
\end{table}

\begin{figure}[ht!]
\centering
\includegraphics[scale=0.25]{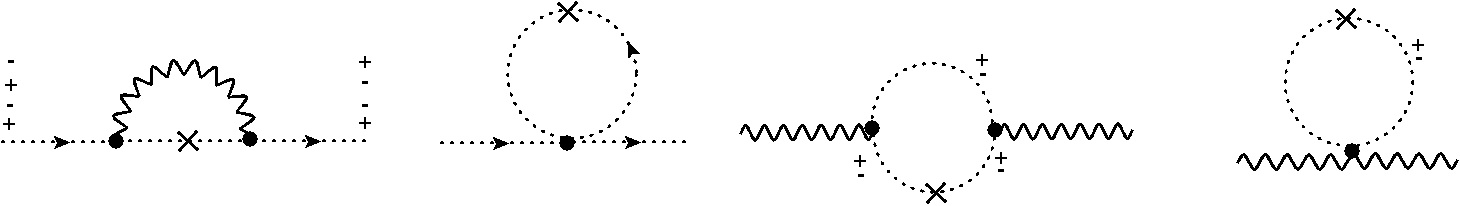}
\caption{One-loop Feynman diagrams leading to the renormalization of dimension-2 squark bilinear operators and to the potential mixing coefficients among themselves and/or with gluon bilinears. A cross corresponds to squark operators. A wavy (dotted) line represents gluons (squarks). Squark lines are further marked with a $+$($-$) sign, to denote an $A_+ \, (A_-)$ field. A squark line arrow entering (exiting) a vertex denotes a $A_+$ ($A_+^{\dagger}$) field; the opposite is true for $A_-$ ($A_-^{\dagger}$) fields.
}
\label{mixing1}
\end{figure}

\begin{figure}[ht!]
\centering
\includegraphics[scale=0.25]{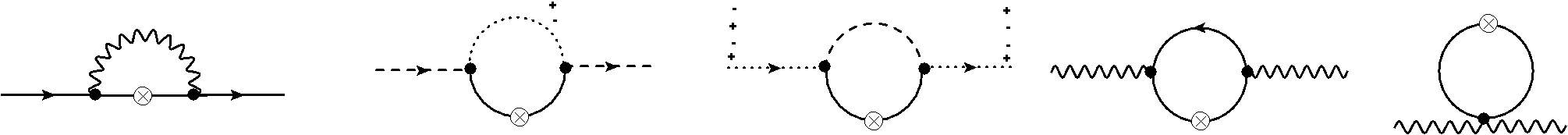}
\caption{One-loop Feynman diagrams leading to the renormalization of dimension-3 quark bilinear operators and to the potential mixing coefficients with gluino, squark and 
gluon bilinears. A circled cross corresponds to quark operators. A wavy (solid) line represents gluons (quarks). A dotted (dashed) line corresponds to squarks (gluinos). Squark lines are further marked with a $+$($-$) sign, to denote an $A_+ \, (A_-)$ field. A squark line arrow entering (exiting) a vertex denotes a $A_+$ ($A_+^{\dagger}$) field; the opposite is true for $A_-$ ($A_-^{\dagger}$) fields.}
\label{mixing2}
\end{figure}

In order to calculate the one-loop mixing coefficients relevant to the squark- and quark-bilinear operators of lowest dimensionality, we must evaluate Feynman diagrams as shown in Figs.~\ref{mixing1} and \ref{mixing2}, respectively. In Figure~\ref{mixing1} we have also included diagrams with external gluons; actually,  since there are no BRST-invariant dimension-2 gluon operators, no mixing is expected to appear in this case, and we use this fact as a check on our perturbative results on the lattice. The diagrams in Figure~\ref{mixing2} lead to the renormalization of dimension-3 quark bilinear operators and to the potential mixing coefficients with gluino, squark and gluon bilinears.

\newpage
\section{Details of the calculation and results}
\label{sec3}

\subsection{Renormalization of quark and squark propagators}
\begin{figure}[ht!]
\centering
\includegraphics[scale=0.3]{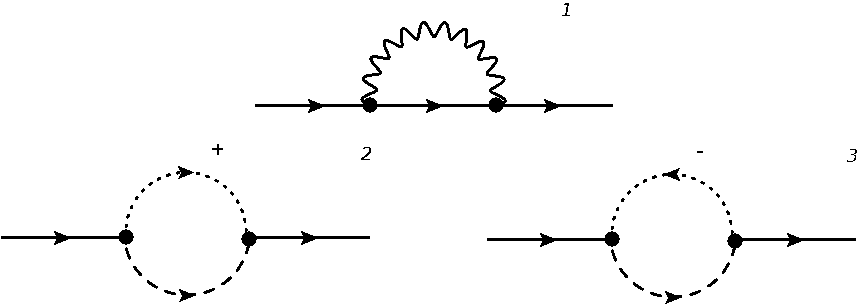}
\caption{One-loop Feynman diagrams contributing to the 2-pt Green's function $\langle \psi(x) \bar \psi(y) \rangle$. A wavy (solid) line represents gluons (quarks). A dotted (dashed) line corresponds to squarks (gluinos). Squark lines are further marked with a $+$($-$) sign, to denote an $A_+ \, (A_-)$ field. A squark line arrow entering (exiting) a vertex denotes a $A_+$ ($A_+^{\dagger}$) field; the opposite is true for $A_-$ ($A_-^{\dagger}$) fields.
  }
\label{quark2pt}
\end{figure}

\begin{figure}[ht!]
\centering
\includegraphics[scale=0.3]{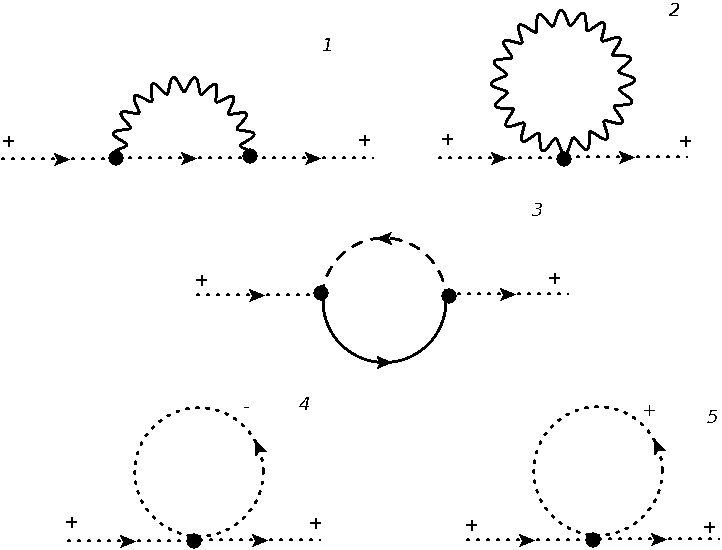}
\caption{One-loop Feynman diagrams contributing to the 2-pt Green's function 
$\langle A_+(x) A_+^{\dagger}(y) \rangle$. The case of $\langle  A_-^{\dagger}(x) A_-(y) \rangle$ is completely analogous. The cases $\langle A_+(x) A_-(y) \rangle$ and $\langle  A_-^{\dagger}(x) A_+^{\dagger}(y) \rangle$ involve only the third Feynman diagram shown above.
  }
\label{squark2pt}
\end{figure}

The purpose of this section is to renormalize the quark and squark fields, as a prerequisite for the renormalization of bilinear composite fields. As a by-product, we also obtain the renormalization factors for the corresponding masses. We use both the dimensional and lattice regularizations of SQCD, in order to calculate the massive quark and squark propagators. Mass and field renormalizations as dictated by renormalization conditions in fact suffice to render finite all the terms in the inverse quark and squark propagators (Eqs. (\ref{Quarkprop}) and (\ref{MIXmatrix}), respectively). The one-loop Feynman diagrams (one-particle irreducible (1PI)) contributing to the quark propagator, $\langle \psi(x) \bar \psi(y) \rangle$,  are shown in Fig.~\ref{quark2pt}, those contributing to the squark propagators, $\langle A_+(x) A^{\dagger}_+(y) \rangle$, $\langle A^{\dagger}_-(x) A_-(y) \rangle$, $\langle A_+(x) A_-(y) \rangle$, $\langle A_-^{\dagger}(x) A^{\dagger}_+(y) \rangle$, in Fig.~\ref{squark2pt}. For our continuum results, we use $\MSbar$ renormalization in the HV ('t Hooft-Veltman) scheme \cite{Hooft:1972} and for completeness we present also the conversion factors to the RI$'$ scheme. In our calculation of the quark propagator the indices carried by all gamma matrices are eventually contracted with the indices of external momenta; thus given that the latter only have 4 (rather than $D$) components, all prescriptions of $\gamma_5$ in $D$-dimensions \cite{Larin, Siegel:1979}, give the same 1-loop results. 

It is convenient to express the squark field components as a doublet: $ A \equiv \left( {\begin{array}{c} A_+ \\ A_-^\dagger \end{array} } \right)$; the mass term then assumes the form $A^\dagger (m_A^\dagger m_A) A$, where the matrix $(m_A^\dagger m_A)$ is hermitian with non-negative eigenvalues, but not necessarily diagonal.

The definitions of the renormalization factors for the matter fields and their masses are:
\bea
\psi^R &=& \sqrt{Z_\psi}\,\psi^B,\\
\label{condS}
A^R &=& \sqrt{Z_{A_\pm}}\,A^B, \\
m_\psi^R &=& Z_{{m}_\psi}\,m_\psi^B, \\ 
{{m^R_{A}}}^\dagger m_A^R &=& Z_{{m}_A}^\dagger {{m^B_{A}}}^\dagger m_A^B Z_{{m}_A},
\label{condMS}
\eea
where $B$ stands for the bare and $R$ for renormalized quantities and $Z_A$, $Z_{{m}_A}$ are $2\times2$ matrices corresponding to the doublet A.
After summing all the continuum Feynman diagrams of Fig.~\ref{quark2pt} the massive inverse quark propagator in dimensional regularization (DR) becomes:
\bea
\langle \tilde \psi^B(q) \tilde{\bar{\psi}}^B(q') \rangle^{DR}_{\rm{inv}} &=& (2\pi)^4 \delta(q-q') \Bigg\{ (i \qslash-m^B_\psi)  +\frac{g^2 C_F}{16 \pi^2} \Bigg[ i \qslash \Bigg(4+\alpha + \frac{2+\alpha}{\epsilon}+(2-\alpha)\frac{m^2}{q^2} +(2+\alpha) \log \left(\frac{\bar \mu^2}{q^2+m^2}\right)\nonumber\\\nonumber  
&&-\left((2-\alpha)\frac{m^4}{q^4}+4\frac{m^2}{q^2}\right)\log \left(1+\frac{q^2}{m^2}\right)\Bigg)\nonumber\\
&&-m \Bigg(4+2 \alpha+\frac{3+\alpha}{\epsilon}+(3+\alpha) \log \left(\frac{\bar \mu^2}{q^2+m^2}\right) 
-(3+\alpha)\frac{m^2}{q^2} \log \left(1+\frac{q^2}{m^2}\right)\Bigg)\Bigg]\Bigg\}\nonumber\\
&=&(2\pi)^4 \delta(q-q') \Big( (i \qslash-m) -\Sigma_\psi^B\Big),
\label{Quarkprop}
\eea
where $C_F=(N_c^2-1)/(2\,N_c)$ is the quadratic Casimir operator in the fundamental representation, $q$ is the external momentum in the Feynman diagrams, and $\bar\mu$ is the $\MSbar$ renormalization scale. For one-loop calculations, the distinction between $m^R$ and $m^B$ is inessential in many cases; we will simply use $m$ in those cases. We have also imposed that the renormalized masses of one flavor for quark and squark be the same. Note also that a Kronecker delta for color indices is understood in Eqs.~(\ref{Quarkprop}) and (\ref{MIXmatrix}). The results for the DR renormalization factors in the $\MSbar$ scheme are:
\bea
Z_\psi^{DR,\MSbar} &=& 1 + \frac{g^2\,C_F}{16\,\pi^2} \frac{1}{\epsilon}\left(2 + \alpha \right)\\
Z_{{m}_\psi}^{DR,\MSbar} &=&  1 + \frac{g^2\,C_F}{16\,\pi^2} \frac{1}{\epsilon} 
\eea
Using our results for the quark propagator, we can compute also the multiplicative renormalization function of the quark field and mass in the RI$'$ renormalization scheme ($Z_\psi^{DR,RI'}$ and $Z_{{m}_\psi}^{DR,RI'}$). In order to find $Z_\psi^{DR,RI'}$, we use the renormalization condition:
\be
\left[ \left(Z_\psi^{DR,RI'}\right)^{-1}i \qslash  -\Sigma_\psi^B\Big|_{{\rm terms}\, \propto \qslash}\right]_{q_\rho=\bar \mu_\rho}   = i \qslash |_{q_\rho= \bar \mu_\rho}\,,
\label{codPsi}
\ee
where $\bar \mu$ is the renormalization scale 4-vector, $\Sigma_\psi^B$ is the quark self energy that we compute up to ${\cal O}(g^2)$. We note that the renormalization scale $\bar \mu$ appearing in $Z_\psi^{DR,RI'}$ need not coincide with the scale used in the $\MSbar$ scale. The RI$'$ counterpart for the multiplicative renormalization of the mass, $Z_{{m}_\psi}^{DR,RI'}$, can be extracted from:
\be
\left[- \left(Z_\psi^{DR,RI'}\right)^{-1} \left(Z_{{m}_\psi}^{DR,RI'}\right)^{-1} m^R  -\Sigma_\psi^B\Big|_{{\rm terms} \,\propto 1}\right]_{q_\rho=\bar \mu_\rho}   = - m^R.
\label{codMPsi}
\ee
The expressions for the aforementioned renormalization factors are:
\bea
Z_\psi^{DR,RI'} &=& 1 + \frac{g^2\,C_F}{16\,\pi^2} \Bigg[ \frac{1}{\epsilon}\left(2 + \alpha \right) +4 + \alpha + (2+\alpha)\log \left(\frac{\bar \mu^2}{\bar \mu^2+m^2}\right) +\frac{m^2}{\bar \mu^2}\left(2-\alpha - 4\log \left(1+\frac{\bar \mu^2}{m^2}\right) \right)\\\nonumber
&&- \frac{m^4}{\bar \mu^4}(2-\alpha) \log \left(1+\frac{\bar \mu^2}{m^2}\right) \Bigg]\\
Z_{{m}_\psi}^{DR,RI'} &=&  1 + \frac{g^2\,C_F}{16\,\pi^2} \Bigg[\frac{1}{\epsilon} + \alpha + \log \left(\frac{\bar \mu^2}{\bar \mu^2+m^2}\right) -\frac{m^2}{\bar \mu^2}\left(2-\alpha - (1-\alpha)\log \left(1+\frac{\bar \mu^2}{m^2}\right) \right)\\\nonumber
&&+\frac{m^4}{\bar \mu^4}(2-\alpha) \log \left(1+\frac{\bar \mu^2}{m^2}\right) 
\Bigg]
\eea
Notice that the expression for $Z_{{m}_\psi}^{DR,RI'}$ is not gauge independent; this was expected given that the renormalization condition relies on a gauge-variant Green's function. The ratio between the $\MSbar$ and RI$'$ renormalization factors give the corresponding conversions factors. 
\bea
C_\psi^{\MSbar,RI'} &=&Z_\psi^{DR,\MSbar}/Z_{{m}_\psi}^{DR,RI'}\\
C_{{m}_\psi}^{\MSbar,RI'}&=&Z_{{m}_\psi}^{DR,\MSbar}/Z_{{m}_\psi}^{DR,RI'}
\eea
Being regularization independent, these same conversion factors can then be also used in the lattice regularization (L). Note also that continuum regularizations forbid the additive renormalization of the mass, so it is renormalized only multiplicatively, while in the lattice regularization the Lagrangian mass $m^L$ undergoes additive and multiplicative renormalization.

\begin{figure}[ht!]
\centering
\includegraphics[scale=0.3]{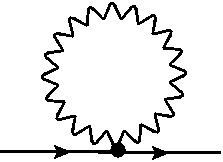}
\caption{One-loop additional lattice Feynman diagram contributing to the 2-pt Green's function $\langle \psi(x) \bar \psi(y) \rangle$.
 }
\label{gluonTquark}
\end{figure}

We next compute the lattice expression of the massive quark propagator to one loop. We take into account the gluon tadpole diagram which has no analog in the continuum (see Fig. \ref{gluonTquark}). Our result is given by:
\bea
\langle \tilde \psi^B(q) \tilde{\bar{\psi}}^B(q') \rangle^{L}_{\rm{inv}} &=& (2\pi)^4 \delta(q-q') \Bigg\{ (i \qslash-m^L_\psi)  +\frac{g^2 C_F}{16 \pi^2} \Bigg[ i \qslash \Bigg(-12.80254+ 4.79201\alpha +(2-\alpha)\frac{m^2}{q^2}\nonumber\\\nonumber  
&&-(2+\alpha) \log \left(a^2(m^2+q^2)\right) - \left( (2-\alpha)\frac{m^4}{q^4}+4\,\frac{m^2}{q^2}\right)\log \left(1+\frac{q^2}{m^2}\right)\Bigg) \nonumber\\\nonumber 
&&-m\Bigg( \left( 0.30799+ 5.7920 \alpha\right) -(3+\alpha) \log \left(a^2(m^2+q^2)\right) 
-(3+\alpha)\frac{m^2}{q^2} \log \left(1+\frac{q^2}{m^2}\right)\Bigg)\nonumber\\
&+&\frac{1}{a} 51.4347 r\Bigg]\Bigg\}\nonumber\\
&=&(2\pi)^4 \delta(q-q') \Big( (i \qslash-m) -\Sigma_\psi^{B,L}\Big)
\label{Quarkproplatt}
\eea

In Eq.~(\ref{Quarkproplatt}), just as in the corresponding equation in the continuum, terms with $\gamma_5$ cancel out at one-loop level. This means that the Majorana components of $\psi$, corresponding to $A_+$ and $A_-$, do not mix under renormalization, unlike the case of the squark fields themselves, see Eqs.~(\ref{MIXmatrix}) and (\ref{MIXmatrixL}). The renormalization factors of the quark fields as well as of the quark mass in the $\MSbar$ scheme and on the lattice are extracted from the subtraction of the renormalized self-energy contributions, which were computed in the continuum, from the bare quark lattice self-energy, as required by the renormalization condition. The critical mass, $m_{crit.}^{quark} \equiv m^L - m^B$, can also be read off Eq.~(\ref{Quarkproplatt}); thus, we find: 

\bea
Z_\psi^{L,\MSbar} &=& 1 + \frac{g^2\,C_F}{16\,\pi^2} \left(-16.8025 + 3.79201\alpha - (2+ \alpha)\log \left(a^2 \bar \mu^2 \right)  \right)\\
Z_{{m}_\psi}^{L,\MSbar} &=&  1 + \frac{g^2\,C_F}{16\,\pi^2}\left(13.1105 -  \log \left(a^2 \bar \mu^2 \right) \right)\\
m^{quark}_{crit.} &=& \frac{g^2\,C_F}{16\,\pi^2} \frac{1}{a}  51.4347  \,r
\eea

We now turn to the one-loop corrections to the squark propagator. Given the mixing of squarks in the HV scheme, our results are written in matrix notation:
\bea
\hspace{-0.5cm}\langle \tilde A^B(q) {\tilde A}^{B\,\dagger}(q') \rangle^{HV}_{\rm{inv}}&=& (2\pi)^4 \delta(q-q')\Bigg[ (q^2 +m^2)\begin{pmatrix} 1 & 0\\ 0 & 1 \end{pmatrix} + q^2 \frac{g^2\,C_F}{16\,\pi^2} \left( \frac{16}{3} + \frac{1+\alpha}{\epsilon}+ (1+\alpha) \log\left(\frac{\bar\mu^2}{q^2+m^2}\right)\right)\begin{pmatrix} 1 & 0 \\ 0 & 1 \end{pmatrix} \nonumber\\
&&+m^2 \frac{g^2\,C_F}{16\,\pi^2} \left(18 + \frac{7+\alpha}{\epsilon} +(7+\alpha) \log\left(\frac{\bar\mu^2}{m^2}\right)- 8 \log \left(1+\frac{q^2}{m^2}\right) - (7-\alpha)\frac{m^2}{q^2}\log \left(1+\frac{q^2}{m^2}\right)\right)\begin{pmatrix} 1 & 0 \\ 0 & 1 \end{pmatrix} \nonumber\\
&&-  \frac{g^2\,C_F}{16\,\pi^2} \left(\frac{4}{3} q^2 + 4 m^2 \right)\begin{pmatrix} 0 & 1\\ 1 & 0 \end{pmatrix}\Bigg]\nonumber\\
&\equiv&(2\pi)^4 \delta(q-q')\left[ (q^2 +m^2) \openone - \Sigma^{B}_A \right].
\label{MIXmatrix}
\eea
where $A^B$ is a 2-component column which contains the bare squark fields.

Starting from  Eq.~(\ref{MIXmatrix}), one requires the elimination of the pole part and determines the $\MSbar$ renormalized 2-pt Green's function. Thus, one arrives at the expressions below for the squark field and mass renormalization respectively.
\bea
Z_A^{DR,\MSbar} &=& \openone \left[1+ \frac{g^2\,C_F}{16\,\pi^2} \frac{1}{\epsilon}\left(1 + \alpha \right)\right]\\
Z_{{m}_A}^{DR,\MSbar} &=&  \openone \left[1 + \frac{g^2\,C_F}{16\,\pi^2} \frac{3}{\epsilon} \right]
\eea
In the naive Dimensional Regularization (NDR) presecription, in which $\gamma_5$ anticommutes with all $\gamma_\mu$ matrices ($\mu = 1,\cdots, D$) \cite{Furman:2003},  the nondiagonal elements in Eq.~(\ref{MIXmatrix}) vanish and the expression for the bare squark propagator in NDR scheme up to one-loop is:
\bea
\label{MIXmatrixNDR}
\hspace{-0.7cm}\langle \tilde A^B(q) {\tilde A}^{B\,\dagger}(q') \rangle^{NDR}_{\rm{inv}}&=& (2\pi)^4 \delta(q-q')\Bigg[ (q^2 +m^2)\begin{pmatrix} 1 & 0\\ 0 & 1 \end{pmatrix} + q^2 \frac{g^2\,C_F}{16\,\pi^2} \left(4 + \frac{1+\alpha}{\epsilon}+ (1+\alpha) \log\left(\frac{\bar\mu^2}{q^2+m^2}\right)\right)\begin{pmatrix} 1 & 0 \\ 0 & 1 \end{pmatrix} \nonumber\\
&+&m^2 \frac{g^2\,C_F}{16\,\pi^2} \left(14 + \frac{7+\alpha}{\epsilon} +(7+\alpha) \log\left(\frac{\bar\mu^2}{m^2}\right)- 8 \log \left(1+\frac{q^2}{m^2}\right) - (7-\alpha)\frac{m^2}{q^2}\log \left(1+\frac{q^2}{m^2}\right)\right)\begin{pmatrix} 1 & 0 \\ 0 & 1 \end{pmatrix} \Bigg]. \nonumber\\
\eea

In computing the conversion factors between $\MSbar$ and $\MSbar$NDR schemes, we cannot simply set $Z^\MSbar/Z^{\MSbar NDR}=1$ because the two regularization actually lead to different renormalization schemes. Instead, we use the definitions of the conversion factors:
\bea
A^\MSbar &=& C_A^{\MSbar,\MSbar\rm{NDR}}A^{\MSbar\rm{NDR}}\\
m^{{2}^\MSbar}_A &=& C_{{m}_A}^{\MSbar,\MSbar\rm{NDR}} m_A^{{2}^{\MSbar\rm{NDR}}} \left(C_{{m}_A}^{\MSbar,\MSbar\rm{NDR}}\right)^\dagger
\eea
which lead to the following values:
\bea
C_A^{\MSbar,\MSbar\rm{NDR}} &=&\openone + \frac{g^2\,C_F}{16\,\pi^2} \frac{4}{3}\begin{pmatrix} -1 & \phantom{-}1\\ \phantom{-}1 &- 1 \end{pmatrix}\\
C_{{m}_A}^{\MSbar,\MSbar\rm{NDR}}&=&\openone  + \frac{g^2\,C_F}{16\,\pi^2} \frac{4}{3}\begin{pmatrix} -1 & \phantom{-}1\\ \phantom{-}1 &- 1 \end{pmatrix}
\eea
Turning now to $RI'$ renormalization, there is a certain amount of freedom in defining it; an essential property which one would like to require is amenability to nonperturbative treatment. As will become clear below, when we discuss the lattice regularization, a natural definition satisfying this requirement is as follows:

\bea
\label{condmA}
\langle \tilde A^{RI'}(q) {\tilde A}^{RI'\,\dagger}(q') \rangle_{\rm{inv}}\Big|_{q^2=0}&=&{m^{R}}^2\openone\\
\langle \tilde A^{RI'}(q) {\tilde A}^{RI'\,\dagger}(q') \rangle_{\rm{inv}}\Big|_{q^2=\bar \mu^2}&=&(q^2 + {m^{R}}^2)|_{q^2=\bar \mu^2}\openone
\label{condA}
\eea
where $\langle \tilde A^{RI'}(q) {\tilde A}^{RI'\,\dagger}(q') \rangle_{\rm{inv}}$ is the RI$'$ renormalized inverse squark propagator which is connected to the bare one through:
\be
\langle \tilde A^{RI'}(q) {\tilde A}^{RI'\,\dagger}(q') \rangle_{\rm{inv}} = \left(Z_A^{DR,RI'}\right)^{1/2} \langle \tilde A^B(q) {\tilde A}^{B\,\dagger}(q') \rangle^{DR}_{\rm{inv}} \left(Z_A^{DR,RI'}\right)^{1/2},
\label{RIS}
\ee
and similarly the renormalized mass $m^{RI'}$ is related to the bare mass $m^B$ through: 
\be
{m^{RI'}}^2 = {Z_m^{L,RI'}}^\dagger {m^B}^2 Z_m^{L,RI'}
\label{RISmass}
\ee
Finally, $\bar \mu$ is the RI$'$ renormalization scale 4-vector. In Eq.~(\ref{condA}) the rhs is the tree level inverse squark propagator. 

Using the renormalization conditions of Eq.~(\ref{condmA})-(\ref{condA}) the multiplicative renomalization in the RI$'$ scheme for the squark field and mass can be determined:
\bea
Z_A^{DR,RI'} &=&  \openone + \frac{g^2\,C_F}{16\,\pi^2} \Bigg[\Bigg( \frac{1+\alpha}{\epsilon}+\frac{16}{3} + (7-\alpha)\frac{m^2}{\bar \mu ^2} + (1+\alpha) \log\left(\frac{\bar\mu^2}{m^2}\right)- (1+\alpha)\log \left(1+\frac{\bar \mu ^2}{m^2}\right)\nonumber\\
&-&8 \frac{m^2}{\bar \mu ^2}\log \left(1+\frac{\bar \mu ^2}{m^2}\right) - (7-\alpha)\frac{m^4}{\bar \mu^4}\log \left(1+\frac{\bar \mu ^2}{m^2}\right) \Bigg)\begin{pmatrix} 1 & 0 \\ 0 & 1 \end{pmatrix} -\frac{4}{3}\begin{pmatrix} 0 & 1 \\ 1 & 0 \end{pmatrix}\Bigg]\\
Z_{{m}_A}^{DR,RI'} &=& \openone + \frac{g^2\,C_F}{16\,\pi^2} \Bigg[\Bigg( \frac{3}{\epsilon}+\frac{17}{6} + \frac{\alpha}{2} - (7-\alpha)\frac{m^2}{2\bar \mu ^2} + 3 \log\left(\frac{\bar\mu^2}{m^2}\right)+\frac{1}{2}(1+\alpha)\log \left(1+\frac{\bar \mu ^2}{m^2}\right)\nonumber\\
&+&4 \frac{m^2}{\bar \mu ^2}\log \left(1+\frac{\bar \mu ^2}{m^2}\right) + \frac{1}{2} (7-\alpha)\frac{m^4}{\bar \mu^4}\log \left(1+\frac{\bar \mu ^2}{m^2}\right) \Bigg)\begin{pmatrix} 1 & 0 \\ 0 & 1 \end{pmatrix} -\frac{4}{3}\begin{pmatrix} 0 & 1 \\ 1 & 0 \end{pmatrix}\Bigg]
\eea

We now proceed with the lattice massive squark propagator, for which we find:
\bea
\hspace{-0.9cm}\langle \tilde A^B(q) {\tilde A}^{B\,\dagger}(q') \rangle^{L}_{\rm{inv}}&=& (2\pi)^4 \delta(q-q')\Bigg[ (q^2 +{m^L_A}^2)\begin{pmatrix} 1 & 0\\ 0 & 1 \end{pmatrix} \nonumber\\
&&-q^2\frac{g^2\,C_F}{16\,\pi^2} \left(11.0173 - 3.7920 \alpha + (1+\alpha)\left[ \log\left(1+\frac{q^2}{m^2}\right)  +  \log(a^2 m^2) \right]\right)\begin{pmatrix} 1 & 0\\ 0 & 1 \end{pmatrix} \nonumber\\
&&-m^2 \frac{g^2\,C_F}{16\,\pi^2} \left(12.2403 - 3.7920 \alpha +(7+\alpha) \log\left(a^2 m^2\right)- 8 \log \left(1+\frac{q^2}{m^2}\right) - (7-\alpha)\frac{m^2}{q^2}\log \left(1+\frac{q^2}{m^2}\right)\right)\begin{pmatrix} 1 & 0 \\ 0 & 1 \end{pmatrix} \nonumber\\
&&+\frac{g^2\,C_F}{16\,\pi^2}\left( -23.8430\, \frac{1}{a}\,r\, m + 65.3930 \,\frac{1}{a^2} \right)\begin{pmatrix} 1 & 0 \\ 0 & 1 \end{pmatrix} \nonumber\\
&&-  \frac{g^2\,C_F}{16\,\pi^2} \left(1.0087 q^2 + 7.6390 m^2 -8.9275 \,\frac{1}{a}\,r\, m -75.4031\, \frac{1}{a^2}\right)\begin{pmatrix} 0 & 1\\ 1 & 0 \end{pmatrix}\Bigg]\nonumber\\
&\equiv&(2\pi)^4 \delta(q-q')\left[ (q^2 +m^2) \openone - \Sigma^{B,L}_A \right].
\label{MIXmatrixL}
\eea
In the above equation the quantities proportional to $1/a$ and to $1/a^2$ constitute a critical mass term $m^{2\,squark}_{crit.}$ which induces an additive renormalization of the squark mass. 
\be
m^{2\,squark}_{crit.}=- \frac{g^2\,C_F}{16\,\pi^2} \left( \frac{1}{a}\,r\, m \begin{pmatrix} -23.8430 & 8.9275 \\ 8.9275 & -23.8430
\end{pmatrix} + \frac{1}{a^2} \begin{pmatrix}65.3930 & 75.4031 \\ 75.4031 & 65.3930\end{pmatrix} \right).
\ee
Substituting Eq.~(\ref{condS}) and Eq.~(\ref{condMS}) into Eq.~(\ref{MIXmatrixL}), and requiring agreement with the finite parts of Eq.~(\ref{MIXmatrix}) in the $a\to 0$ limit, we find:

\be
Z_A^{L,\MSbar} = \openone + \frac{g^2\,C_F}{16\,\pi^2}\left[ \left(-16.3507 + 3.79201 \alpha - (1+\alpha) \log\left(a^2 \bar \mu^2\right)\right)\begin{pmatrix} 1 & 0 \\ 0 & 1 \end{pmatrix} +0.32464 \begin{pmatrix} 0 & 1\\ 1 & 0 \end{pmatrix} \right]
\ee
and
\be
Z_{{m}_A}^{L,\MSbar} = \openone - \frac{g^2\,C_F}{16\,\pi^2}\left[ \left(6.94486 + 3 \log\left(a^2 \bar \mu^2\right)\right)\begin{pmatrix} 1 & 0 \\ 0 & 1 \end{pmatrix} +1.98181 \begin{pmatrix} 0 & 1\\ 1 & 0 \end{pmatrix} \right]
\ee

The existence of the critical mass, in the expression of the squark propagator, can be used to calibrate Monte Carlo simulations and to tune the Lagrangian mass. Given that the critical mass is a $2 \times 2$ symmetric matrix of the form: $\begin{pmatrix} a & b \\ b & a \end{pmatrix}$, there are two parameters ($a$, $b$) to be calibrated. Our perturbative results can be used as a starting point for this calibration; $m^B$ is then the difference between the Lagrangian mass and the critical mass. 

Given that the critical masses are power divergent, perturbation theory is expected to provide only a ballpark estimate at best \cite{Catterall:2011, Catterall:2014}. Therefore, non-perturbative estimates are of importance. In order to determine the critical mass non-pertubatively one should find appropriate Ward identities (see, e.g.,~\cite{Zhestkov:2001hx}) and enforce them by appropriate calibration. On the other hand, one may use $(A_+^\dagger A_-^\dagger - A_-A_+)$ whose has as one of its superpartners the $\bar \psi \gamma_5 \psi$. Thus, in calculating correlation functions of these operators, one may tune the Lagrangian mass in such a way that these operators create a massless state from the vacuum, just like the pion, and thus one can determine the critical mass. This can be done, assuming chiral symmetry is spontaneously broken while SUSY is not.

We also propose a general recipe for tuning the lagragian mass non-perturbatively. The quantity we need to know is the renormalized mass $m^R$ in the $\MSbar$ scheme. Actually, even when we use $RI'$ renormalization, we will express our Green's functions in terms of $m^\MSbar$; this is by analogy with the treatment of the renormalized coupling constant in $RI'$. The renormalized mass can be calculated as follows: Substituting Eqs.~(\ref{RIS}) and (\ref{RISmass}) into Eqs.~(\ref{condmA}) and (\ref{condA}) (where the bare Green's functions are non-perturbatively determined via lattice numerical simulations), we obtain 8 conditions for the 8 unknown matrix elements of $Z_A$ and ${Z_m}_A$. Upon subtracting Eq.~(\ref{condmA}) from Eq.~(\ref{condA}), for any given choice of $m^L_A$, we determine $Z_A$, and  subsequently from Eq.~(\ref{condmA}) we determine ${Z_m}_A$. Having determined $Z_A$ and ${Z_m}_A$ non-perturbatively in this manner, we thus arrive at a definite value for $m^R$. In order to achieve a desired value for $m^R$, we may then tune the Lagrangian mass.

\newpage

\subsection{Mixing of bilinear operators}

We compute both in the continnum and on the lattice the matrix elements for quark and squark bilinear operators. From these matrix elements we provide the renormalization of the quark bilinears and the mixing coefficients with gluino bilinear operators, as well as with operators made of gluon and of squark fields \cite{Taniguchi:2000, Vladikas:2002}. In addition, we present the renormalization of dimension-2 squark operators: To one-loop, there is no mixing of these operators among themselves; further, we check that there is no mixing with dimension-2 gluon bilinears either.

\subsubsection{Renormalization of squark bilinear operators}

In the following, we first calculate the case of squark bilinear operators in the continuum, where we regularize the theory in $D$ dimensions ($D=4-2\,\epsilon$). The squark bilinears,  ${\cal O}^A$, having dimension 2 can mix in principle with other dimension-2 operators. This entails studying the 2-pt Green's functions of ${\cal O}^A$ with external squark and gluon fields. 
There are four squark operators which we denote as $O^A_{\pm\,\pm}$ :
\bea
{\cal O}_{+\,+}^A(x) &=& A_+^{\dagger}(x)\, A_+(x) \\
{\cal O}_{+\,-}^A(x) &=& A_+^{\dagger}(x)\, A_-^{\dagger}(x)\\
{\cal O}_{-\,+}^A(x) &=& A_-(x)\, A_+(x)\\
{\cal O}_{-\,-}^A(x) &=& A_-(x)\, A_-^{\dagger}(x)
\eea
Given that all the quantities which we set out to calculate are $x$-independent, we will often apply an integration (or a summation on the lattice) over $x$ for convenience.

We note that some of these operators may develop a vacuum expectaion value in the flavor singlet case ($\frac{1}{N_f} \sum_f A_f^\dagger A_f$ ), leading to a mixing with the unit operator. Thus, the tree level vacuum expectation values of ${\cal O}_{+\,+}^A$ and  ${\cal O}_{+\,-}^A$ are:
\be
\langle {\cal O}_{+\,+}^A(x) \rangle^{DR} = \langle {\cal O}_{-\,-}^A(x) \rangle^{DR} = 
\frac{N_c}{(4\pi)^{D/2}}m^{D-2} \Gamma(1-D/2).
\ee
The vacuum expectation values of ${\cal O}_{-\,+}^A$ and ${\cal O}_{+\,-}^A$ vanish. In order to eliminate mixing with the unit operator, all operators will be taken to be normal ordered; thus, their vacuum expectation values will be subtracted from their definition.

To one-loop order, the expression of the 2-pt Green's functions of the squark bilinear operators, with external squark fields, are given by the sum of the first two Feynman diagrams shown in Fig.~\ref{mixing1}. In the flavor non-singlet case, the two squark propagators appearing to the left and right of the operator insertion contain different masses. However, it is sufficient to evaluate the diagrams in the mass-degenerate case, for the following reasons: a) The pole parts, which determine renormalization and mixing coefficients in DR, are mass independent in all cases, b) The difference between lattice and continuum bare Green's functions, which gives the corresponding coefficients on the lattice, is also mass independent.

We present the amputated 1PI 2-pt Green's functions with external squark fields in the continuum: 

\bea
\langle \tilde A_+(q) {\cal O}_{+\,+}^A  \tilde A_+^{\dagger}(q') \rangle_\amp^{DR} &=&  \langle \tilde A_-^{\dagger}(q) {\cal O}_{-\,-}^A  \tilde A_-(q') \rangle_\amp^{DR}= \nonumber \\ 
&&\hspace{-4cm}(2\pi)^4 \delta(q-q')\Bigg\{ 1 +  \frac{g^2\,C_F}{16\,\pi^2} \bigg[ \frac{\alpha-1}{\epsilon} -2(\alpha-3) + (\alpha-1) \log \left(\frac{\bar{\mu}^2}{m^2} \right) + 2(\alpha-3) \frac{m^2}{q^2}\log \left(1+\frac{q^2}{m^2} \right) \bigg]\Bigg\}
\label{plusplus}
\eea

\bea
\langle \tilde A_+(q) {\cal O}_{+\,-}^A   \tilde A_-(q') \rangle_\amp^{DR} &=& \langle \tilde A_-^{\dagger}(q) {\cal O}_{-\,+}^A   \tilde A_+^{\dagger}(q') \rangle_\amp^{DR} = \nonumber\\  
&&\hspace{-4cm}(2\pi)^4 \delta(q-q')\Bigg\{ 1 +  \frac{g^2\,C_F}{16\,\pi^2} \bigg[ \frac{\alpha+1}{\epsilon} -2(\alpha-3) + (\alpha+1) \log \left(\frac{\bar{\mu}^2}{m^2} \right) + 2(\alpha-3) \frac{m^2}{q^2}\log \left(1+\frac{q^2}{m^2} \right) \bigg]\Bigg\}
\label{plusminus}
\eea

In order to determine the renormalization factors for the squark bilinear operators, ${\cal O}_{\pm\,\pm}^A(x)$, we use the renormalization conditions in Eq~(\ref{squarkBilinearCondDR}), requiring that the lhs be finite. 

\be
\langle \tilde A^R_a {{\cal O}_{b\,c}^A}^R \tilde {A^R_d}^{\dagger} \rangle_\amp = 
{(Z_A)}^{-1/2}_{a\,a'} Z_{b c b' c'} \langle  \tilde A^B_{a'} {{\cal O}_{b'\,c'}^A}^B \tilde {A^B_{d'}}^{\dagger} \rangle_\amp (Z_A)^{-1/2}_{d'\,d},  \qquad {{\cal O}_{a\,b}^A}^R = Z_{a\,b\,a'\,b'} {{\cal O}_{a'\,b'}^A}^B,
\label{squarkBilinearCondDR}
\ee
where the indices ${a, b, c, d, a', b', c', d'}$ can take the values $+$ and $-$ , as shown in Eqs.~(\ref{plusplus}) and (\ref{plusminus}); a summation over any repeated index is understood. 

At one-loop order, using these conditions, we find that:
\be
Z_{+\,+\,+\,+}^{DR,\MSbar}=Z_{-\,-\,-\,-}^{DR,\MSbar}= 1 + \frac{g^2\,C_F}{16\,\pi^2} \frac{2}{\epsilon}
\ee
\be
Z_{+\,-\,+\,-}^{DR,\MSbar}=Z_{-\,+\,-\,+}^{DR,\MSbar}= 1
\ee
All other components of $Z_{a\,b\,a'\,b'}$ vanish. Thus in the 't Hooft-Veltman regularization scheme, with $\MSbar$ renormalization, the operators ${\cal O}_{+\,-}^A$ and ${\cal O}_{-\,+}^A$ receive no corrections up to one loop; furthermore, there is no mixing between any of these operators.

In order to check that there is no mixing with other Lorentz scalar dimension-2 gluon operators, we calculate the diagrams of the squark bilinear operators with external gluons, as shown in the last two diagrams of Fig.~\ref{mixing1}. By studying the corresponding Green's functions we find that indeed this case receives no mixing, and thus flavor singlet squark operators cannot mix with the gluon bilinear $u_\mu u^\mu$. In particular, there was a cancelation of the pole part in these diagrams leading to the expected result. Indeed, the 2-pt Green's functions of these operators with external gluons turn out to be finite and  transverse. Their numerical expressions are:
\bea
\langle \tilde u_\mu^\alpha(q) {\cal O}_{+\,+}^A  \tilde u_\nu^\beta(q') \rangle_\amp^{DR} &=&  \langle \tilde u_\mu^\alpha(q) \tilde {\cal O}_{-\,-}^A  \tilde u_\nu^\beta(q') \rangle_\amp^{DR}= \nonumber \\ 
&&\hspace{-4cm}-(2\pi)^4 \delta(q+q') \delta^{\alpha \beta} \frac{g^2}{16\,\pi^2} \left(q^2 \delta_{\mu \nu} - q_{\mu} q_{\nu}\right) \bigg[ \frac{2}{q^2} - 2 \frac{\sqrt{4m^2+q^2}}{q^3} \log \left(\frac{q+\sqrt{4m^2+q^2}}{2m} \right) \bigg]
\label{SquarkOpDRGluon}
\eea

Since $u_\mu u^\mu$ is not type I, II or III, the lattice regulator also does not allow mixing with this operator. We check the above, calculating the same quantities on the lattice. Our results coincide with those of the continuum, Eq.~(\ref{SquarkOpDRGluon}). 
\bigskip

Let us now turn to the lattice contributions. The vacuum expectation values of ${\cal O}_{+\,+}^A(x)$ and ${\cal O}_{+\,-}^A(x)$ on the lattice are, in the flavor singlet case:
\be
\langle {\cal O}_{+\,+}^A(x) \rangle = \langle {\cal O}_{-\,-}^A(x) \rangle = N_c \int_{-\pi/a}^{\pi/a}\frac{d^4p}{(2\pi)^4}\frac{1}{\frac{4}{a^2} \sum_\mu \sin^2(a p_\mu/2)+m^2} = \frac{N_c}{a^2} \left( 0.154933390 -\frac{a^2\,m^2}{\pi^2} (0.29950-\frac{1}{16}\log(a^2\,m^2))\right)
\ee

Regarding the lattice Green's functions of squark bilinear operators with external squarks, our results are: 
\bea
\langle \tilde A_+(q) {\cal O}_{+\,+}^A  \tilde A_+^{\dagger}(q') \rangle_\amp^{L} &=&  \langle \tilde A_-^{\dagger}(q) {\cal O}_{-\,-}^A  \tilde A_-(q') \rangle_\amp^{L}= \nonumber \\ 
&&\hspace{-5cm}(2\pi)^4 \delta(q-q')\Bigg\{ 1 +  \frac{g^2\,C_F}{16\,\pi^2} \bigg[
4 + 1.7920 (\alpha-1)-  (\alpha-1) \log \left(m^2\,a^2 \right) + 2(\alpha-3) \frac{m^2}{q^2}\log \left(1+\frac{q^2}{m^2} \right) \bigg]\Bigg\}
\eea

\bea
\langle \tilde A_+(q) {\cal O}_{+\,-}^A   \tilde A_-(q') \rangle_\amp^{L} &=& \langle \tilde A_-^{\dagger}(q) {\cal O}_{-\,+}^A   \tilde A_+^{\dagger}(q') \rangle_\amp^{L} = \nonumber\\  
&&\hspace{-5cm}(2\pi)^4 \delta(q-q')\Bigg\{ 1 +  \frac{g^2\,C_F}{16\,\pi^2} \bigg[ 
8 + 1.7920 (1+\alpha)- (1+\alpha) \log \left(m^2 a^2 \right) + 2(\alpha-3) \frac{m^2}{q^2}\log \left(1+\frac{q^2}{m^2} \right)
 \bigg]\Bigg\}
\eea

Requiring that the above bare lattice Green's functions, upon renormalization, lead to the same expressions as the continuum ones (Eq.~(\ref{squarkBilinearCondDR})), we arrive at the following lattice renormalization factors: 

\be
Z_{+\,+\,+\,+}^{L,\MSbar}=Z_{-\,-\,-\,-}^{L,\MSbar}= 1 - \frac{g^2\,C_F}{16\,\pi^2} \left(12.5586 + 2 \log(a^2 \bar\mu^2) \right)
\label{plusplusPP}
\ee
\be
Z_{+\,-\,+\,-}^{L,\MSbar}=Z_{-\,+\,-\,+}^{L,\MSbar}= 1 - \frac{g^2\,C_F}{16\,\pi^2}\left(20.1462\right)
\label{plusminusPM}
\ee
Once again, even though our calculations were performed with flavors of equal masses, the results Eqs.~(\ref{plusplusPP})-(\ref{plusminusPM}) are valid also in the mass-nondegenerate case.

\subsubsection{Renormalization of quark bilinear operators}

In this subsection we calculate the 2-pt Green's functions of the quantities ${\cal O}_i^\psi$, using both dimensional regularization (DR) and lattice regularization (L). As shown in Fig.~\ref{mixing2}, there are nine (1+2+4+1+1) one-loop Feynman diagrams which involve different external fields.  These diagrams correspond to seven Green's functions which we must compute in order to identify the renormalization and the mixing patterns of ${\cal O}_i^\psi$.
 
From the first diagram in Fig.~\ref{mixing2} we calculate the renormalization factors $Z_{i}$ of the quark bilinear operators ${\cal{O}}_i^\psi$. The other diagrams contribute to the mixing coefficients $z^\la_{i}$, $z^{\pm\,\pm}_{i}$, $z^{\pm D\pm}_{i}$, $z^{m\pm \pm}_{i}$, $z^u_{i}$, with gluino, squark (involving zero or one derivatives, or one power of the mass) and gluon bilinear operators, respectively. The expressions relevant for the mixing of each quark bilinear assume the following forms:
\bea
\label{scalar}
{{\cal O}_S^\psi}^R &=& Z_S {{\cal O}_S^\psi}^B + z^\la_S {{\cal O}_S^\la}^B + z^{+\,+}_S ({{\cal O}_{+\,+}^A}^B+{{\cal O}_{-\,-}^A}^B) + z^{+\,-}_S ({{\cal O}_{+\,-}^A}^B + {{\cal O}_{-\,+}^A}^B)  \\ \nonumber
&& + z^{m+\,+}_S (m_f + m_{f'})({{\cal O}_{+\,+}^A}^B+{{\cal O}_{-\,-}^A}^B) + z^{m+\,-}_S (m_f + m_{f'}) ({{\cal O}_{+\,-}^A}^B + {{\cal O}_{-\,+}^A}^B)   \\
{{\cal O}_P^\psi}^R &=& Z_P {{\cal O}_P^\psi}^B + z^\la_P {{\cal O}_P^\la}^B + 
z^{+\,-}_P ({{\cal O}_{+\,-}^A}^B - {{\cal O}_{-\,+}^A}^B) 
+z^{m+\,+}_P (m_f - m_{f'}) ({{\cal O}_{+\,+}^A}^B-{{\cal O}_{-\,-}^A}^B)\\ \nonumber
&& +z^{m+\,-}_P (m_f + m_{f'}) ({{\cal O}_{+\,-}^A}^B - {{\cal O}_{-\,+}^A}^B)  \\
{{\cal O}_{V,\mu}^\psi}^R &=& Z_{V} {{\cal O}_{V,\mu}^\psi}^B + z^\la_V {{\cal O}_{V,\mu}^\la}^B + z^{+D+}_V (A_+^{\dagger} D_\mu A_+ + A_- D_\mu  A_-^{\dagger}) \\\nonumber
&&+ z_V^{+D-}(A_+^{\dagger} D_\mu A_-^{\dagger} + A_- D_\mu A_+) +z^u_V u_\mu \partial_\nu u_\nu
\\
{{\cal O}_{AV,\mu}^\psi}^R &=& Z_{AV} {{\cal O}_{AV,\mu}^\psi}^B + z^\la_{AV} {{\cal O}_{AV,\mu}^\la}^B + z_{AV}^{+D+}(A_+^{\dagger} D_\mu A_+^{\dagger} - A_- D_\mu A_-)+z^u_{AV} \epsilon_{\mu\,\nu\,\rho\,\sigma}u_\nu \partial_\rho u_\sigma\\
{{\cal O}_T^\psi}^R &=& Z_T {{\cal O}_T^\psi}^B + z^\la_T {{\cal O}_T^\la}^B
\label{tensor}
\eea
On the rhs of Eqs.~(\ref{scalar})-(\ref{tensor}) there appear all operators that can possibly mix with those on the lhs; the tree-level Green's functions of these mixing operators naturally show up in the results for the one-loop Green's functions of the quark operators, thus allowing us to deduce the corresponding mixing coefficients. In the case of flavor nonsinglet operators, $z^\la_i$ and $z^u_i$ automatically vanish. Note that for the scalar and pseudoscalar operators, the presence of the mixing with linear combinations of ${\cal O}_{-\,+}^A$ and ${\cal O}_{+\,-}^A$ comes from parity and charge conjugation. We have also introduced at this point the mixing coefficients $z^{\pm D \pm}_i$ which correspond to the vector and axial vector operators shown in Table \ref{tb:non-singlet} with one covariant derivative in their definition; once again, parity and charge conjugation dictate their relevant linear combinations. The axial vector quark operator yields an expression related to the axial anomaly. The latter stems from the last diagram of Fig.~\ref{mixing2}, which involves non-supersymmetric particles; thus, it must reproduce the equivalent result in QCD. The lattice discretization should give the correct axial anomaly term in the continuum limit. Lastly, the tensor quark operator can only mix with the gluino osions are our results in $DR$:

\bea
\langle  \tilde \psi^B(q) {{\cal O}_S^\psi}^B \tilde{\bar{\psi}}^B(q') \rangle^{DR}_{\rm{amp}} &=&  (2 \pi)^4 \delta(q-q') \Bigg\{\openone \Bigg[ 1 + \frac{g^2\,C_F}{16\,\pi^2} \Bigg( \frac{3+\alpha}{\epsilon} + 4 + 2 \alpha  +  (3 + \alpha) \log\left(\frac{\bar\mu^2}{m^2} \right) \\\nonumber
&-& (3+\alpha)\left(1+3\frac{m^2}{q^2}  \right) \log\left(1+\frac{q^2}{m^2} \right)  \Bigg) \Bigg] +4 i \frac{g^2\,C_F}{16\,\pi^2} \alpha \qslash \left(\frac{m}{q^2}+\frac{m^3}{q^4}\log\left(1+\frac{q^2}{m^2} \right)  \right)\Bigg\}\\
\langle  \tilde \psi^B(q) {{\cal O}_P^\psi}^B \tilde{\bar{\psi}}^B(q') \rangle^{DR}_{\rm{amp}}&=& (2 \pi)^4 \delta(q-q') \gamma_5 \Bigg[ 1 + \frac{g^2\,C_F}{16\,\pi^2} \Bigg( \frac{3+\alpha}{\epsilon} + 4 + 2 \alpha  +  (3 + \alpha) \log\left(\frac{\bar\mu^2}{m^2} \right) \\\nonumber
&-& (3+\alpha)\left(1+\frac{m^2}{q^2}\right)  \log\left(1+\frac{q^2}{m^2} \right)  \Bigg) \Bigg]\\
\langle  \tilde \psi^B(q) {{\cal O}_V^\psi}^B \tilde{\bar{\psi}}^B(q') \rangle^{DR}_{\rm{amp}}&=& (2 \pi)^4 \delta(q-q') \Bigg\{\gamma_\mu \Bigg[ 1 + \frac{g^2\,C_F}{16\,\pi^2}  \alpha \Bigg( \frac{1}{\epsilon} + 1  +  \log\left(\frac{\bar\mu^2}{m^2}\right) - \log\left(1+\frac{q^2}{m^2} \right) \\\nonumber
&& -\frac{m^2}{q^2} +\frac{m^4}{q^4} \log\left(1+\frac{q^2}{m^2} \right) \Bigg)   \Bigg] \\ \nonumber
&& + q_\mu \qslash  \frac{g^2\,C_F}{16\,\pi^2} \alpha \left( 4\frac{m^2}{q^2} - 2  \frac{1}{q^2} - 4 \frac{m^4}{q^6}\log\left(1+\frac{q^2}{m^2} \right)  \right) \\\nonumber
&& - i \frac{g^2\,C_F}{16\,\pi^2} q_\mu (6+2\alpha)\left( \frac{m}{q^2}-\frac{m^3}{q^4} \log\left(1+\frac{q^2}{m^2} \right) \right)\Bigg\}\\
\langle  \tilde \psi^B(q) {{\cal O}_{AV}^\psi}^B \tilde{\bar{\psi}}^B(q') \rangle^{DR}_{\rm{amp}} &=& (2 \pi)^4 \delta(q-q') \Bigg\{ \gamma_5 \gamma_\mu \Bigg[ 1 + \frac{g^2\,C_F}{16\,\pi^2}   \Bigg( \frac{\alpha}{\epsilon} + \alpha + \alpha \log\left(\frac{\bar\mu^2}{m^2} \right) - \alpha\log\left(1+\frac{q^2}{m^2} \right) \\\nonumber
&& -(2-\alpha)\frac{m^2}{q^2}-2(1+\alpha)\frac{m^2}{q^2}\log\left(1+\frac{q^2}{m^2} \right)+(2-\alpha)\frac{m^4}{q^4}\log\left(1+\frac{q^2}{m^2} \right)\Bigg)\Bigg]\\\nonumber
&& + i  \frac{g^2\,C_F}{16\,\pi^2}\gamma_5 q_\mu  \left(2(1-\alpha)\frac{m}{q^2} - 2(1-\alpha)\frac{m^3}{q^4} \log\left(1+\frac{q^2}{m^2} \right)  \right) \\\nonumber
&& - i \frac{g^2\,C_F}{16\,\pi^2} \gamma_5 \gamma_\mu \qslash \left( 2(1-\alpha)\frac{m}{q^2}- 2(1-\alpha)\frac{m^3}{q^4} \log\left(1+\frac{q^2}{m^2} \right) \right)\\\nonumber
&&-\frac{g^2\,C_F}{16\,\pi^2} \gamma_5 q_\mu \qslash \Bigg(2\alpha\frac{1}{q^2}-4(2-\alpha) \frac{m^2}{q^4}+4(1-\alpha)\frac{m^2}{q^4}\log\left(1+\frac{q^2}{m^2} \right)\\\nonumber
&& +4(2-\alpha) \frac{m^4}{q^6} \log\left(1+\frac{q^2}{m^2} \right)\Bigg) \Bigg\}\\
\langle  \tilde \psi^B(q) {{\cal O}_T^\psi}^B \tilde{\bar{\psi}}^B(q') \rangle^{DR}_{\rm{amp}}&=& (2 \pi)^4 \delta(q-q') \Bigg\{\frac{1}{2}[\gamma_\mu, \gamma_\nu] \Bigg[ 1 + \frac{g^2\,C_F}{16\,\pi^2}  \left(\alpha - 1 \right)\Bigg( \frac{1}{\epsilon} +  2\frac{m^2}{q^2}+ \log\left(\frac{\bar\mu^2}{m^2}\right) \\\nonumber
&&- \left( 1 + \frac{m^2}{q^2} +  2\frac{m^4}{q^4}  \right)\log\left(1+\frac{q^2}{m^2}  \right) \Bigg) \Bigg]\\\nonumber
&&+ 4 i\frac{g^2\,C_F}{16\,\pi^2} (\gamma_\mu q_\nu - \gamma_\nu q_\mu)\left( \frac{m}{q^2}- \frac{m^3}{q^4} \log\left(1+\frac{q^2}{m^2}  \right) \right)\\\nonumber
&&-4 i\frac{1}{2}[\gamma_\mu, \gamma_\nu]\qslash \left( \frac{m}{q^2}- \frac{m^3}{q^4} \log\left(1+\frac{q^2}{m^2}  \right) \right)\\\nonumber
&&+\frac{g^2\,C_F}{16\,\pi^2}(\gamma_\mu q_\nu - \gamma_\nu q_\mu)\qslash\left(4(1-\alpha)\frac{m^2}{q^4}-2(1-\alpha)\left(\frac{m^2}{q^4}+2\frac{m^4}{q^6}\right)\log\left(1+\frac{q^2}{m^2}  \right)\right)\Bigg\}
\eea

It is worth noting that the pole terms are analogous to the tree level Green's functions of the operators. There appear additional, finite contributions with tensor structures which are distinct from those at tree level. For the case of the vector and axial vector operators, such tensor structures appear even in the limit of zero mass. The determination of the renormalization factors can be achieved by imposing the renormalization conditions of Eq.~(\ref{quarkBilinearCondDR}), and demanding the lhs to be finite; as usual, $Z_i$ may only contain pole terms beyond tree level in the $\MSbar$ scheme.  
\be
\langle \tilde \psi^R {{\cal O}_i^\psi}^R \tilde {\bar \psi}^R \rangle_\amp = Z_i Z_\psi^{-1} \langle \tilde \psi^B {{\cal O}_i^\psi}^B \tilde {\bar \psi}^B \rangle_\amp
\label{quarkBilinearCondDR}
\ee
Therefore, the continuum renormalization factors are: 
\bea
{Z_S}^{DR,\MSbar} &=& 1 - \frac{g^2\,C_F}{16\,\pi^2} \frac{1}{\epsilon} \\
{Z_P}^{DR,\MSbar} &=& 1 - \frac{g^2\,C_F}{16\,\pi^2} \frac{1}{\epsilon} \\
{Z_V}^{DR,\MSbar} &=& 1 + \frac{g^2\,C_F}{16\,\pi^2} \frac{2}{\epsilon} \\
{Z_{AV}}^{DR,\MSbar} &=& 1 + \frac{g^2\,C_F}{16\,\pi^2} \frac{2}{\epsilon} \\
{Z_T}^{DR,\MSbar} &=& 1 + \frac{g^2\,C_F}{16\,\pi^2} \frac{3}{\epsilon}.
\eea
In what follows the $\MSbar$ scheme actually refers to HV, and consequently it provides the well known result for the axial current, ABJ anomaly. From the above we also keep the $\MSbar$ renormalized Green's functions which are essential ingredients in order to extract the lattice renormalization factors. We also note that the results for $Z_{{m}_\psi}$ and $Z_S$ are related by $Z_{{m}_\psi} = Z^{- 1}_S$ as expected.

The next step in our renormalization procedure is to calculate the mixing coefficients. We concentrate on the Green's functions of ${{\cal O}_i^\psi}$ with external gluino, squark and gluon fields. Taking as an example the scalar quark operator with external squark fields, the renormalized Green's function will be given by:

\be
\langle \tilde A^R {{\cal O}_S^\psi}^R {\tilde A}^{R \dagger} \rangle_\amp = Z_S Z_A^{-1/2} \langle \tilde A^B {{\cal O}_S^\psi}^B \tilde {A^B}^\dagger \rangle_\amp Z_A^{-1/2} - \sum_{a,b=+,-}z_S^{a\,b} Z_A^{-1/2} \langle \tilde A^B {\cal O}_{a\,b}^A \tilde {A^B}^\dagger \rangle_\amp^{\rm{tree}}Z_A^{-1/2}
\label{quarkBilMIXING}
\ee
($z_S^{-\,-} \equiv z_S^{+\,+}$, $z_S^-+ \equiv z_S^{+\,-}$, cf. Eq.~(\ref{scalar})).
Similarly, taking into account the potential mixing with ${{\cal O}_i^\la}$, which appears only in the flavor singlet case, and the corresponding tree-level Green's functions we can determine $z^\la_i$. The expressions we obtain for $\langle  \la^B {{\cal O}_i^\psi}^B \bar \la^B \rangle^{DR}_{\rm{amp}}$ are shown here (A Kronecker delta is understood over the color indices of the external gluino fields, which are left implicit):
\bea
\label{DRscalarL}
\langle  \tilde \la^B(q) {{\cal O}_S^\psi}^B \tilde {\bar \la}^B(q') \rangle^{DR}_{\rm{amp}} &=& 
(2 \pi)^4 \delta(q-q') \frac{g^2\,}{16\,\pi^2} \Bigg[ i \qslash \left(\frac{2m}{q^2} - \frac{8m^3}{q^3\,\sqrt{4m^2+q^2}}\log \left(\frac{q+\sqrt{4m^2+q^2}}{2m} \right) \right)\Bigg]\\
\langle  \tilde \la^B(q) {{\cal O}_P^\psi}^B \tilde {\bar \la}^B(q') \rangle^{DR}_{\rm{amp}}&=& 0\\
\langle  \tilde \la^B(q) {{\cal O}_V^\psi}^B \tilde {\bar \la}^B(q') \rangle^{DR}_{\rm{amp}}&=& (2 \pi)^4 \delta(q-q')   \frac{g^2\,}{16\,\pi^2} \Bigg[ \gamma_\mu \left( 1 +\frac{1}{2\epsilon} + \frac{1}{2}\log \left(\frac{\bar{\mu}^2}{m^2} \right) - \frac{4m^2+q^2}{q\,\sqrt{4m^2+q^2}}\log \left(\frac{q+\sqrt{4m^2+q^2}}{2m} \right) \right) \nonumber\\
&+& \qslash q_\mu \left(-\frac{1}{q^2}+ \frac{4m^2}{q^3\,\sqrt{4m^2+q^2}}\log \left(\frac{q+\sqrt{4m^2+q^2}}{2m} \right)\right) \Bigg]\\
\langle  \tilde \la^B(q) {{\cal O}_{AV}^\psi}^B \tilde {\bar \la}^B(q') \rangle^{DR}_{\rm{amp}}&=& (2 \pi)^4 \delta(q-q')   \frac{g^2\,}{16\,\pi^2} \Bigg[ \gamma_5 \gamma_\mu\left( -2 -\frac{1}{2\epsilon} - \frac{1}{2}\log \left(\frac{\bar{\mu}^2}{m^2} \right) +  \frac{8m^2+q^2}{q\,\sqrt{4m^2+q^2}}\log \left(\frac{q+\sqrt{4m^2+q^2}}{2m} \right) \right) \nonumber\\
&+& \gamma_5 \qslash q_\mu \left(\frac{1}{q^2}- \frac{4m^2}{q^3\,\sqrt{4m^2+q^2}}\log \left(\frac{q+\sqrt{4m^2+q^2}}{2m} \right)\right) \Bigg]\\
\langle  \tilde \la^B(q) {{\cal O}_T^\psi}^B \tilde {\bar \la}^B(q') \rangle^{DR}_{\rm{amp}}&=& (2 \pi)^4 \delta(q-q')   \frac{g^2\,}{16\,\pi^2} \Bigg[ \frac{1}{2}[\gamma_\mu, \gamma_\nu] i \qslash \left(\frac{2m}{q^2} - \frac{8m^3}{q^3\,\sqrt{4m^2+q^2}}\log \left(\frac{q+\sqrt{4m^2+q^2}}{2m} \right) \right) \nonumber\\
&-& i (\gamma_\mu q_\nu - \gamma_\nu q_\mu) \left(\frac{2m}{q^2}- \frac{8m^3}{q^3\,\sqrt{4m^2+q^2}}\log \left(\frac{q+\sqrt{4m^2+q^2}}{2m} \right)\right) \Bigg]
\label{DRtensorL}
\eea

There are 4 possibilities corresponding to Green's functions of each quark bilinear with external squarks, according to the choice of squark components, $A_+$ or $A_-$. For the sake of a concise presentation, the corresponding expressions are shown below in the mass-degenerate case, with the necessary exception of a factor $\Delta m \equiv m_f - m_{f'}$, which is related to the mixing with $\Delta m (A_+^{\dagger} A_+-A_- A_-^{\dagger})$: Indeed, these expressions are sufficient for  the extraction of all renormalizaton coefficients on the lattice, see Eqs.~(\ref{squarkSPPM})-(\ref{squarkSPPP}).

\bea
\label{DRscalarS}
\langle \tilde A^B_+(q) {{\cal O}_S^\psi}^B  \tilde {A_+^{\dagger}}^B (q') \rangle^{DR}_{\rm{amp}} &=&  \langle \tilde{A_-^{\dagger}}^B (q) {{\cal O}_S^\psi}^B  \tilde A^B_-(q') \rangle^{DR}_{\rm{amp}} =-(2 \pi)^4 \delta(q-q') \frac{g^2\,C_F}{16\,\pi^2} \Bigg[24m + 16m \Bigg(\frac{1}{\epsilon} \\\nonumber
&+& \log \left(\frac{\bar{\mu}^2}{m^2+q^2} \right) - \frac{m^2}{q^2} \log \left(1+ \frac{q^2}{m^2} \right)\Bigg) \Bigg]\\
\langle \tilde A^B_+(q) {{\cal O}_S^\psi}^B  \tilde A^B_-(q') \rangle^{DR}_{\rm{amp}} &=& \langle \tilde {A_-^{\dagger}}^B(q) {{\cal O}_S^\psi}^B \tilde {A_+^{\dagger}}^B(q') \rangle^{DR}_{\rm{amp}} = (2 \pi)^4 \delta(q-q') \frac{g^2\,C_F}{16\,\pi^2} (8 m) \\
\langle  \tilde A^B_+(q) {{\cal O}_P^\psi}^B \tilde {A_+^{\dagger}}^B (q') \rangle^{DR}_{\rm{amp}} &=& -\langle \tilde {A_-^{\dagger}}^B(q) {{\cal O}_P^\psi}^B  \tilde A^B_-(q') \rangle^{DR}_{\rm{amp}} = -(2 \pi)^4 \delta(q-q') \frac{g^2\,C_F}{16\,\pi^2}  \Delta m \Bigg[\frac{8}{\epsilon} \\\nonumber
&+&12+ 8\log \left(\frac{\bar{\mu}^2}{m^2+q^2} \right) - 8\frac{m^2}{q^2} \log \left(1+ \frac{q^2}{m^2} \right)\Bigg]\\
\langle  \tilde A^B_+(q) {{\cal O}_P^\psi}^B  \tilde A^B_-(q') \rangle^{DR}_{\rm{amp}} &=& -\langle  \tilde {A_-^{\dagger}}^B(q) {{\cal O}_P^\psi}^B \tilde {A_+^{\dagger}}^B(q') \rangle^{DR}_{\rm{amp}} =  (2 \pi)^4 \delta(q-q') \frac{g^2\,C_F}{16\,\pi^2} (8 m) \\
\langle  \tilde A^B_+(q) {{\cal O}_V^\psi}^B \tilde {A_+^{\dagger}}^B (q') \rangle^{DR}_{\rm{amp}} &=& \langle \tilde {A_-^{\dagger}}^B(q) {{\cal O}_V^\psi}^B  \tilde A^B_-(q') \rangle^{DR}_{\rm{amp}} =-(2 \pi)^4 \delta(q-q') \frac{g^2\,C_F}{16\,\pi^2} i\,q_\mu \Bigg[\frac{32}{3}+\frac{8}{\epsilon} \\\nonumber
&-&8\frac{m^2}{q^2} + 8 \log \left(\frac{\bar{\mu}^2}{m^2+q^2} \right) +8\frac{m^4}{q^4} \log \left(1+ \frac{q^2}{m^2} \right)\Bigg] \\
\langle \tilde A^B_+(q) {{\cal O}_V^\psi}^B  \tilde A^B_-(q') \rangle^{DR}_{\rm{amp}} &=& \langle \tilde {A_-^{\dagger}}^B(q) {{\cal O}_V^\psi}^B \tilde  {A_+^{\dagger}}^B(q') \rangle^{DR}_{\rm{amp}}=  (2 \pi)^4 \delta(q-q') \frac{g^2\,C_F}{16\,\pi^2} i\,q_\mu\left(\frac{8}{3} \right)\\
\langle  \tilde A^B_+(q) {{\cal O}_{AV}^\psi}^B \tilde  {A_+^{\dagger}}^B (q') \rangle^{DR}_{\rm{amp}} &=&  -\langle \tilde {A_-^{\dagger}}^B(q) {{\cal O}_{AV}^\psi}^B  \tilde A^B_-(q') \rangle^{DR}_{\rm{amp}}  =(2 \pi)^4 \delta(q-q') \frac{g^2\,C_F}{16\,\pi^2} i\,q_\mu \Bigg[16+\frac{8}{\epsilon} \\\nonumber
&+& 8\frac{m^2}{q^2} + 8 \log \left(\frac{\bar{\mu}^2}{m^2+q^2} \right)  - 16\frac{m^2}{q^2} \log \left(1+ \frac{q^2}{m^2} \right)-8\frac{m^4}{q^4} \log \left(1+ \frac{q^2}{m^2} \right)\Bigg] \\
\langle  \tilde A^B_+(q) {{\cal O}_{AV}^\psi}^B  \tilde A^B_-(q') \rangle^{DR}_{\rm{amp}} &=&  \langle \tilde {A_-^{\dagger}}^B(q) {{\cal O}_{AV}^\psi}^B \tilde  {A_+^{\dagger}}^B(q') \rangle^{DR}_{\rm{amp}}= 0\\
\langle \tilde  A^B_+(q) {{\cal O}_{T}^\psi}^B \tilde {A_+^{\dagger}}^B (q') \rangle^{DR}_{\rm{amp}} &=&  \langle \tilde {A_-^{\dagger}}^B(q) {{\cal O}_{T}^\psi}^B  \tilde A^B_-(q') \rangle^{DR}_{\rm{amp}}=0\\
\langle  \tilde A^B_+(q) {{\cal O}_{T}^\psi}^B  \tilde A^B_-(q') \rangle^{DR}_{\rm{amp}} &=&  \langle \tilde {A_-^{\dagger}}^B(q) {{\cal O}_{T}^\psi}^B \tilde  {A_+^{\dagger}}^B(q') \rangle^{DR}_{\rm{amp}}=0
\label{DRtensorS}
\eea

Lastly, we compute the gluon matrix elements of the quark bilinears:
\bea
\label{DRscalarG}
\langle \tilde u_\sigma^B(q) {{\cal O}_S^\psi}^B \tilde u_\nu^B(q') \rangle^{DR}_{\rm{amp}}&=& (2 \pi)^4 \delta(q+q')  \frac{g^2\,}{16\,\pi^2}  \Bigg[\left(\delta_{\sigma\,\nu}-\frac{q_\sigma q_\nu}{q^2}  \right)\times\\\nonumber
&&\phantom{(2 \pi)^4 \delta(q+q')  \frac{g^2\,}{16\,\pi^2} }\Bigg(8m -32 \frac{m^3}{q\,\sqrt{4m^2+q^2}}\log \left(\frac{q+\sqrt{4m^2+q^2}}{2m} \right) \Bigg)  \Bigg]\\
\langle  \tilde u_\sigma^B(q) {{\cal O}_{AV}^\psi}^B \tilde u_\nu^B(q') \rangle^{DR}_{\rm{amp}}&=& (2 \pi)^4 \delta(q+q')  \frac{g^2\,}{16\,\pi^2} \Bigg[\epsilon_{\sigma\,\nu\,\mu\,\rho} i q_\rho   \Bigg(-4 +\frac{16m^2}{q\,\sqrt{4m^2+q^2}}\log \left(\frac{q+\sqrt{4m^2+q^2}}{2m} \right)   \Bigg)\Bigg]\\
\langle  \tilde u_\sigma^B(q) {{\cal O}_V^\psi}^B \tilde u_\nu^B(q') \rangle^{DR}_{\rm{amp}}&=& \langle  \tilde u_\sigma^B(q) {{\cal O}_{P}^\psi}^B \tilde u_\nu^B(q') \rangle^{DR}_{\rm{amp}}=\langle  \tilde u_\sigma^B(q) {{\cal O}_{T}^\psi}^B \tilde u_\rho^B(q') \rangle^{DR}_{\rm{amp}}=0
\label{DRtensorG}
\eea
Following the example of Eq.~(\ref{quarkBilMIXING}), and using the definition of $Z_i$ and the renormalization of each field, all mixing coefficients $z^\la_i$, $z^{\pm\,\pm}_i$ ($z_{i}^{\pm D \pm}$, $z^{m\pm\,\pm}_i$),  $z^u_i$ can be derived from Eqs.~(\ref{DRscalarL})-(\ref{DRtensorG}), respectively:

\bea
\hspace{-1cm} z^\la_S=z^\la_P=z^\la_T=0&,&\,z^\la_V=\frac{g^2\,}{16\,\pi^2}\frac{1}{\epsilon},\,z^\la_{AV}=-\frac{g^2\,}{16\,\pi^2}\frac{1}{\epsilon} \\
\,z^{+\,+}_S&=&  z^{+\,-}_S= z_P^{+\,-}=0\\
z^{m+\,+}_S&=&  -\frac{g^2\,}{16\,\pi^2}\frac{8}{\epsilon},\, z^{m+\,-}_S=0\\
z^{m+\,+}_P&=& -\frac{g^2\,}{16\,\pi^2}\frac{8}{\epsilon},\, z^{m+\,-}_P=0\\
z^{+D+}_{V}&=&-\frac{g^2\,}{16\,\pi^2}\frac{8}{\epsilon},\,z^{+D-}_V=0\\
z^{+D+}_{AV}&=&\frac{g^2\,}{16\,\pi^2}\frac{8}{\epsilon}\\ 
z^{u}_V&=&z^{u}_{AV}=0
\eea


Having calculated the above quantities in the continuum, we proceed with the computation of the lattice Green's functions in order to extract the renormalization and mixing coefficients \cite{Konishi} of to the quark bilinears. Since on the lattice some symmetries are broken, e.g. chiral symmetry, more mixings arise \cite{Rattazzi}; they contain inverse powers of the lattice spacing, and thus they require an independent non-perturbative determination. In this work we have evaluated pertubatively all the renormalizations and mixing coefficients at one loop. 

The computation of the 2-pt bare Green's functions of ${\cal O}_i^\psi$ on the lattice are the most demanding part of the present work. The algebraic expressions involved were split into two parts: a) Terms that can be evaluated in the $a \to 0$ limit: Such terms exhibit a very complicated dependence on the external momentum, even for zero masses. This dependence constitutes a part of the regularization independent renormalized Green's functions. b) All remaining terms: These are divergent as $a \to 0$, however their dependence on $q$, $m$ is necessarily polynomial. The lattice introduces additional contributions, which are finite and polynomial in $q$ and $m$. Our computations were performed in a covariant gauge, with arbitrary value of the gauge parameter $\alpha$. Given that some of the operators which mix with ${\cal O}_i^\psi$ contain powers of the masses, we have kept these masses different from zero throughout the computation. 

Renormalizability of the theory implies that the difference between the one-loop renormalized and bare Green's functions must only consist of expressions which are polynomial in $q$, $m$; in this way, the rhs of Eq.~(\ref{quarkBilMIXING}) can be rendered equal to the corresponding lhs, by an appropriate definition of the renormalization factors and mixing coefficients. Indeed, we have checked explicitly the polynomial character of the above differences. This check is quite nontrivial, given the very complex dependence of the initial expressions on the momenta $q$. Both renormalized and bare functions have the same tensorial form, but the bare ones have additional contributions. Each tensorial structure will provide an equation (cf. Eq.~(\ref{quarkBilMIXING})); these equations can be solved indepedently for each mixing coefficient. 

As mentioned, we employ the HV scheme which is more useful for comparison with experimental determinations and phenomenological estimates. Certain results for the mixing coefficients will depend also on the lattice spacing. For completeness, we write all relevant Green's functions, shown in Eqs.~(\ref{qSqL})-(\ref{qTqL}) and (\ref{gluinoL})-(\ref{gluonL}). All results presented in this section are computed for nonzero masses and momentum. Additionally, we should note that the errors on our lattice expressions are smaller than the last shown digit.

\bea
\label{qSqL}
\langle  \tilde \psi^B(q) {{\cal O}_S^\psi}^B \tilde{\bar{\psi}}^B(q') \rangle^{L}_{\rm{amp}} &=&  \langle  \tilde \psi^\MSbar(q) {{\cal O}_S^\psi}^\MSbar \tilde{\bar{\psi}}^\MSbar(q') \rangle_{\rm{amp}} -\\\nonumber 
&&  (2 \pi)^4 \delta(q-q') \frac{g^2\,C_F}{16\,\pi^2} \Big(3.69200  - 3.79201 \alpha 
+ (3+ \alpha) \log(a^2 \bar\mu^2) \Big)\\
\langle  \tilde \psi^B(q) {{\cal O}_P^\psi}^B \tilde{\bar{\psi}}^B(q') \rangle^{L}_{\rm{amp}} &=&  \langle  \tilde \psi^\MSbar(q) {{\cal O}_P^\psi}^\MSbar \tilde{\bar{\psi}}^\MSbar(q') \rangle_{\rm{amp}} -\\\nonumber  
&&(2 \pi)^4 \delta(q-q') \frac{g^2\,C_F}{16\,\pi^2} \Big (-5.95103 - 3.79201 \alpha 
+ (3+ \alpha) \log(a^2 \bar\mu^2) \Big)\\
\langle  \tilde \psi^B(q) {{\cal O}_V^\psi}^B \tilde{\bar{\psi}}^B(q') \rangle^{L}_{\rm{amp}} &=&  \langle  \tilde \psi^\MSbar(q) {{\cal O}_V^\psi}^\MSbar \tilde{\bar{\psi}}^\MSbar(q') \rangle_{\rm{amp}} -\\\nonumber   
&&(2 \pi)^4 \delta(q-q') \frac{g^2\,C_F}{16\,\pi^2} \Big (-3.97338 - 3.79201 \alpha \
+ \alpha \log(a^2 \bar\mu^2) \Big)\\
\langle  \tilde \psi^B(q) {{\cal O}_A^\psi}^B \tilde{\bar{\psi}}^B(q') \rangle^{L}_{\rm{amp}} &=&  \langle  \tilde \psi^\MSbar(q) {{\cal O}_A^\psi}^\MSbar \tilde{\bar{\psi}}^\MSbar(q') \rangle_{\rm{amp}} -\\\nonumber 
&&(2 \pi)^4 \delta(q-q') \frac{g^2\,C_F}{16\,\pi^2} \Big (0.84813 - 3.79201 \alpha  
+ \alpha \log(a^2 \bar\mu^2) \Big)\\
\label{qTqL}
\langle  \tilde \psi^B(q) {{\cal O}_T^\psi}^B \tilde{\bar{\psi}}^B(q') \rangle^{L}_{\rm{amp}} &=&  \langle  \tilde \psi^\MSbar(q) {{\cal O}_T^\psi}^\MSbar \tilde{\bar{\psi}}^\MSbar(q') \rangle_{\rm{amp}} - \\\nonumber  
&&(2 \pi)^4 \delta(q-q') \frac{g^2\,C_F}{16\,\pi^2} \Big (-0.37366 - 3.79201 \alpha 
+ (-1+\alpha) \log(a^2 \bar\mu^2) \Big).
\eea
The $\MSbar$-renormalized quantities on the rhs of the above equations are equal to the corresponding expressions in Eqs.~(\ref{qSqL})-(\ref{qTqL}), with all $1/\epsilon$ poles removed; same applies to Eqs.~(\ref{gluinoL})-(\ref{gluonL}). These results are completely compatible with results in Refs.~\cite{Skouroupathis:2007jd, Skouroupathis:2008mf}.

By combining the lattice expressions with the renormalized Green's functions calculated in the continuum (see Eq.~(\ref{quarkBilinearCondDR})), we find for the renormalization factors: 
\bea
{Z_S}^{L,\MSbar} &=& 1 + \frac{g^2\,C_F}{16\,\pi^2}  \left( -13.1105 + \log(a^2\bar\mu^2)\right)\\
{Z_P}^{L,\MSbar} &=& 1 + \frac{g^2\,C_F}{16\,\pi^2}  \left(-22.7536 + \log(a^2\bar\mu^2)\right)\\
{Z_{V}}^{L,\MSbar} &=& 1 + \frac{g^2\,C_F}{16\,\pi^2}  \left(-20.7759 -2 \log(a^2\bar\mu^2)\right)\\
{Z_{AV}}^{L,\MSbar} &=& 1 + \frac{g^2\,C_F}{16\,\pi^2}  \left(-15.9544 -2 \log(a^2\bar\mu^2)\right)\\
{Z_T}^{L,\MSbar} &=& 1 + \frac{g^2\,C_F}{16\,\pi^2} \left(-17.1762 -3 \log(a^2 \bar\mu^2) \right).
\eea
We note that these factors are all gauge independent, as they should be in the $\MSbar$ scheme. The remaining quantities which we need to compute on the lattice are the mixing coefficients. These can be easily determined by computing the Green's functions corresponding to the last four diagrams of Fig.~\ref{mixing2} in the lattice regularization.  For several of these Green's functions, their difference from the corresponding functions in $DR$ vanishes, in the limit $a \to 0$, with no additional lattice contributions; thus, for the sake of brevity, we list below only those cases in which the difference is nonvanishing. 

\bea
\label{gluinoL}
\langle  \la^B(q) {{\cal O}_V^\psi}^B {\bar \la}^B(q') \rangle^{L}_{\rm{amp}}&=& \langle  \la^{\MSbar}(q) {{\cal O}_V^\psi}^{\MSbar} {\bar \la}^{\MSbar}(q') \rangle _{\rm{amp}}+ \frac{g^2}{16\,\pi^2} \Big (2.24195 - \frac{1}{2} \log\left(a^2 \bar \mu^2 \right) \Big)\\
\langle  \la^B(q) {{\cal O}_{AV}^\psi}^B {\bar \la}^B(q') \rangle^{L}_{\rm{amp}}&=& \langle  \la^{\MSbar}(q) {{\cal O}_{AV}^\psi}^{\MSbar} {\bar \la}^{\MSbar}(q') \rangle _{\rm{amp}} +\frac{g^2}{16\,\pi^2} \Big ( 2.85434 + \frac{1}{2} \log\left(a^2 \bar \mu^2 \right)\Big)\\
\label{squarkSPPM}
\langle  A^B_+(q) {{\cal O}_S^\psi}^B  {A_+^{\dagger}}^B (q') \rangle^{L}_{\rm{amp}} &=& \langle  A^{\MSbar}_+(q) {{\cal O}_S^\psi}^{\MSbar}  {A_+^{\dagger}}^{\MSbar} (q') \rangle _{\rm{amp}} + \frac{g^2\,C_F}{16\,\pi^2} \Big (52.8968 \bar m + 23.8429 r \frac{1}{a} \\\nonumber
&&\phantom{\langle  A^{\MSbar}_+(q) {{\cal O}_S^\psi}^{\MSbar}  {A_+^{\dagger}}^{\MSbar} (q') \rangle _{\rm{amp}}}+ 16 \bar m \log\left(a^2 \bar \mu^2 \right) \Big)\\
\langle  {A_-^{\dagger}}^B(q) {{\cal O}_S^\psi}^B  A^B_-(q') \rangle^{L}_{\rm{amp}}&=& \langle  {A_-^{\dagger}}^{\MSbar}(q) {{\cal O}_S^\psi}^{\MSbar}  A^{\MSbar}_-(q') \rangle _{\rm{amp}} + \frac{g^2\,C_F}{16\,\pi^2} \Big (52.8968 \bar m + 23.8429 r \frac{1}{a} \\\nonumber
&&\phantom{\langle  A^{\MSbar}_+(q) {{\cal O}_S^\psi}^{\MSbar}  {A_+^{\dagger}}^{\MSbar} (q') \rangle _{\rm{amp}}}+ 16 \bar m \log\left(a^2 \bar \mu^2 \right)\Big)\\
\langle  A^B_+(q) {{\cal O}_S^\psi}^B  A^B_-(q') \rangle^{L}_{\rm{amp}} &=& \langle  A^{\MSbar}_+(q) {{\cal O}_S^\psi}^{\MSbar}  A^{\MSbar}_-(q') \rangle _{\rm{amp}} +\frac{g^2\,C_F}{16\,\pi^2}\left(7.2780  \bar m - 8.92745  r  \frac{1}{a} \right)\\
\langle  {A_-^{\dagger}}^B(q) {{\cal O}_S^\psi}^B  {A_+^{\dagger}}^B(q') \rangle^{L}_{\rm{amp}} &=& \langle  {A_-^{\dagger}}^{\MSbar}(q) {{\cal O}_S^\psi}^{\MSbar}  {A_+^{\dagger}}^{\MSbar}(q') \rangle _{\rm{amp}} + \frac{g^2\,C_F}{16\,\pi^2}\left(7.2780  \bar m - 8.92745  r \frac{1}{a} \right)\\
\langle  A^B_+(q) {{\cal O}_P^\psi}^B  {A_+^{\dagger}}^B (q') \rangle^{L}_{\rm{amp}} &=&\langle  A^{\MSbar}_+(q) {{\cal O}_P^\psi}^{\MSbar}  {A_+^{\dagger}}^{\MSbar} (q') \rangle _{\rm{amp}} + \frac{g^2\,C_F}{16\,\pi^2} \Delta m \Big(7.9207 + 8\log\left(a^2 \bar \mu^2 \right)\Big) \\ 
\langle  {A_-^{\dagger}}^B(q) {{\cal O}_P^\psi}^B  A^B_-(q') \rangle^{L}_{\rm{amp}}&=&\langle  {A_-^{\dagger}}^{\MSbar}(q) {{\cal O}_P^\psi}^{\MSbar}  A^{\MSbar}_-(q') \rangle _{\rm{amp}} - \frac{g^2\,C_F}{16\,\pi^2} \Delta m \Big(7.9207 + 8\log\left(a^2 \bar \mu^2 \right)\Big)\\
\langle  A^B_+(q) {{\cal O}_P^\psi}^B  A^B_-(q') \rangle^{L}_{\rm{amp}} &=& \langle  A^{\MSbar}_+(q) {{\cal O}_P^\psi}^{\MSbar}  A^{\MSbar}_-(q') \rangle _{\rm{amp}} - \frac{g^2\,C_F}{16\,\pi^2}\left(29.7772 \bar m + 32.7704 r \frac{1}{a}\right)\\
\label{squarkSPPP}
\langle  {A_-^{\dagger}}^B(q) {{\cal O}_P^\psi}^B  {A_+^{\dagger}}^B(q') \rangle^{L}_{\rm{amp}}&=& \langle  {A_-^{\dagger}}^{\MSbar}(q) {{\cal O}_P^\psi}^{\MSbar}  {A_+^{\dagger}}^{\MSbar}(q') \rangle _{\rm{amp}}+ \frac{g^2\,C_F}{16\,\pi^2}\left(29.7772 \bar m + 32.7704 r \frac{1}{a}\right)\\
\langle  A^B_+(q) {{\cal O}_V^\psi}^B  {A_+^{\dagger}}^B (q') \rangle^{L}_{\rm{amp}} &=& \langle  A^{\MSbar}_+(q) {{\cal O}_V^\psi}^{\MSbar}  {A_+^{\dagger}}^{\MSbar} (q') \rangle _{\rm{amp}} + \frac{g^2\,C_F}{16\,\pi^2} i q_\mu\left (5.6888 + 8\log\left(a^2 \bar \mu^2 \right) \right)\\
\langle  {A_-^{\dagger}}^B(q) {{\cal O}_V^\psi}^B  A^B_-(q') \rangle^{L}_{\rm{amp}}&=&\langle  {A_-^{\dagger}}^{\MSbar}(q) {{\cal O}_V^\psi}^{\MSbar}  A^{\MSbar}_-(q') \rangle _{\rm{amp}}+\frac{g^2\,C_F}{16\,\pi^2} i q_\mu\left (5.6888 + 8\log\left(a^2 \bar \mu^2 \right) \right)\\
\langle  A^B_+(q) {{\cal O}_V^\psi}^B  A^B_-(q') \rangle^{L}_{\rm{amp}} &=&  \langle  A^{\MSbar}_+(q) {{\cal O}_V^\psi}^{\MSbar}  A^{\MSbar}_-(q') \rangle _{\rm{amp}} - \frac{g^2\,C_F}{16\,\pi^2} i q_\mu 0.8693\\
\langle  {A_-^{\dagger}}^B(q) {{\cal O}_V^\psi}^B  {A_+^{\dagger}}^B(q') \rangle^{L}_{\rm{amp}}&=& \langle  {A_-^{\dagger}}^{\MSbar}(q) {{\cal O}_V^\psi}^{\MSbar}  {A_+^{\dagger}}^{\MSbar}(q') \rangle_{\rm{amp}} - \frac{g^2\,C_F}{16\,\pi^2} i q_\mu 0.8693\\
\langle  A^B_+(q) {{\cal O}_{AV}^\psi}^B  {A_+^{\dagger}}^B (q') \rangle^{L}_{\rm{amp}} &=&  \langle  A^{\MSbar}_+(q) {{\cal O}_{AV}^\psi}^{\MSbar}  {A_+^{\dagger}}^{\MSbar} (q') \rangle _{\rm{amp}} - \frac{g^2\,C_F}{16\,\pi^2} i q_\mu\left (14.6168 + 8\log\left(a^2 \bar \mu^2 \right) \right)\\
\langle  {A_-^{\dagger}}^B(q) {{\cal O}_{AV}^\psi}^B  A^B_-(q') \rangle^{L}_{\rm{amp}}&=&\langle  {A_-^{\dagger}}^{\MSbar}(q) {{\cal O}_{AV}^\psi}^{\MSbar}  A^{\MSbar}_-(q') \rangle _{\rm{amp}}+\frac{g^2\,C_F}{16\,\pi^2} i q_\mu\left (14.6168 + 8\log\left(a^2 \bar \mu^2 \right) \right)
%
%
%
%
%
%
%
\label{gluonL}
\eea
where $\bar m \equiv (m_f+m_{f'})/2$.

The lattice one-loop expressions for the mixing coefficients are presented here. By inserting our lattice results into Eq.~(\ref{quarkBilMIXING}) (and similarly for all other relevant Green's functions) we immediately obtain:
\bea
\hspace{-1cm} z^\la_S=z^\la_P=z^\la_T=0&,&\,z^\la_V=\frac{g^2}{16\,\pi^2} \Big (4.4839 - \log\left(a^2 \bar \mu^2 \right) \Big),\,z^\la_{AV}= \frac{g^2}{16\,\pi^2} \Big ( 5.7087 + \log\left(a^2 \bar \mu^2 \right)\Big) \\
\,z^{+\,+}_S&=& 
\frac{g^2\,C_F}{16\,\pi^2} \frac{1}{a} 23.8429 r,\, \,\,
z^{+\,-}_S= 
-\frac{g^2\,C_F}{16\,\pi^2} \frac{1}{a} 8.9274 r\\
z^{m+\,+}_S&=& 
\frac{g^2\,C_F}{16\,\pi^2} \left (26.4484 + 8 \log\left(a^2 \bar \mu^2 \right) \right),\,\,\, z^{m+\,-}_S= 
\frac{g^2\,C_F}{16\,\pi^2}3.6390\\
z^{+\,-}_P&=&
-\frac{g^2\,C_F}{16\,\pi^2} \frac{1}{a} 32.7704 r\\
z^{m+\,+}_P&=& \frac{g^2\,C_F}{16\,\pi^2} \Big(7.9207 + 8\log\left(a^2 \bar \mu^2 \right)\Big),\,\,\,
z^{m+\,-}_P=
-\frac{g^2\,C_F}{16\,\pi^2}14.8886\\
z^{+D+}_V&=&
\frac{g^2\,C_F}{16\,\pi^2}\left(5.6888 + 8\log\left(a^2 \bar \mu^2 \right) \right),\,\,\,
z^{+D-}_V=
 -\frac{g^2\,C_F}{16\,\pi^2} 0.8693\\
z^{+D+}_{AV}&=&
-\frac{g^2\,C_F}{16\,\pi^2} \left (14.6168 + 8\log\left(a^2 \bar \mu^2 \right) \right),\, \,\,z^{u}_V=z^{u}_{AV}=0
\eea
The above perturbative estimates of the renormalization factors $Z_i$ and of the mixing coefficients $z_i$ can be used for the determination of the properly renormalized ${\cal O}_i^\psi$ matrix elements.

\section{Summary of Results and future plans}
\label{summary}

In this paper we have studied the mixing under renormalization for local bilinear operators in SQCD. We have calculated the one-loop renormalization factors and mixing coefficients for local quark operators and dimension-2 squark operators, both in Dimensional Regularization (DR) and on the Lattice (L), in the $\MSbar$ renormalization scheme. In the supersymmetric case more mixings arise as compared to QCD, due to the fact that the SQCD action contains more fields and interactions; indeed, in QCD there is no mixing when one calculates the Green's functions of local quark bilinears \cite{Taniguchi:2000}. As a prerequisite, we have computed the quark and squark inverse propagators and thus we have determined the multiplicative renormalization of these fields and of their masses, as well as the critical values for each mass. One novel aspect of this work is that we use the SQCD action with nonzero masses $m_f$ throughout our computations,in order to avoid infrared divergences, thus there emerge more mixing patterns among operators made of quark, squark, gluino and gluons. 

The renormalization factors and the operator mixings can be determined from the calculation of certain 2-pt Green's functions. We have calculated the Green's functions of the dimension-2 squark operators with external squarks and gluons. The latter shows that there is no mixing between squark and gluon operators and from the former we have extracted the renormalization factors for the dimension-2 squark operators. However, for the quark operators, which are dimension 3, there are more mixing patterns. We have calculated the Green's functions of the quark bilinears with external quarks, squarks, gluinos and gluons and we determined all renormalization factors and mixing cofficients to one loop.

A natural continuation of the present work is to calculate the Green's functions of the above operators up to all orders in the lattice spacing. These extensions are useful in order to construct improved versions of the operators, but also to remove ${\cal O}(g^2\,a^\infty)$ contributions from future non-perturbative data. A further extension of the present work would be to calculate the mixing matrices of extended quark bilinears containing one or more covariant derivatives in their definition. These operators have dimension 4 and thus the list of the mixing operators increases. To this end, one should calculate all 2-pt Green's functions and may also need to calculate certain 3-pt Green's functions of these operators. Such Green's functions provide more detailed information on the structure of bound states in SQCD. It would be also interesting to compute the existing Green's functions up to two loops. Computing higher loops in perturbation theory is a difficult task due to the rapidly increasing number of Feynman diagrams in the supersymmetric case and the appearance of more complicated expressions, as well as due to the more intricate structure of (sub)divergences. Also the differences between flavor-singlet and -nonsinglet operators are more pronounced, because of diagrams with closed fermion loops. Lastly, it would be important to extend our computations to further improved actions with reduced lattice artifacts and reduced symmetry breaking, e.g. the overlap fermion action, as a forerunner to numerical studies using these actions.

\end{document}